\documentclass[12pt]{article}
\usepackage{amsmath,amscd,amsfonts,eucal,latexsym,amssymb,mathrsfs}
\newlength{\dinwidth}
\newlength{\dinmargin}
\numberwithin{equation}{section}
 \newtheorem{Definition}{Definition}[section]
 \newtheorem{Theorem}[Definition]{Theorem}
 \newtheorem{Proposition}[Definition]{Proposition}
 \newtheorem{Lemma}[Definition]{Lemma}
 \newtheorem{Corollary}[Definition]{Corollary}

\newsymbol\rest 1316         

\def\cA{{\cal A}}
\def\cB{{\cal B}}

\def\cF{{\cal F}}

\def\cH{{\cal H}}

\def\cJ{{\cal J}}

\def\cL{{\cal L}}
\def\cM{{\cal M}}
\def\cN{{\cal N}}
\def\cO{{\cal O}}

\def\cS{{\cal S}}

\def\cU{{\cal U}}
\def\cV{{\cal V}}
\def\cW{{\cal W}}

\def\cZ{{\cal Z}}

\def\bC{{\mathbb C}}

\def\bN{{\mathbb N}}
\def\NN{{\mathbb N}}
\def\bR{{\mathbb R}}
\def\RR{{\mathbb R}}

\def\bZ{{\mathbb Z}}

\def\a{\alpha}
\def\b{\beta}
\def\g{\gamma}        \def\G{\Gamma}
\def\d{\delta}        \def\D{\Delta}

\def\l{\lambda}       
\def\m{\mu}

\def\r{\rho}
\def\s{\sigma}

\def\t{\tau}

\def\o{\omega}        \def\O{\Omega}

\def\fA{{\mathfrak A}}

\def\fQ{{\mathfrak Q}}

\def\fT{{\mathfrak T}}

\def\imply{\Rightarrow}

\def\supp{{\text{supp}}}

\def\rrb{\boldsymbol{\varrho}}
\def\ux{\underline{x}}
\def\uxi{\underline{\xi}}

\newcommand{\tU}{\widetilde{U}}
\newsymbol\bt 1202  

\newcommand{\Cin}{C^{\infty}}
\newcommand{\Coin}{C^{\infty}_{0}}

\newcommand{\tst}{\tau_{t}{}_*}
\newcommand{\tss}{\tau_s{}_*}

\newcommand{\hide}[1]{} 

\newcommand{\gb}{{\boldsymbol{g}}}
\newcommand{\hb}{\boldsymbol{h}}
\newcommand{\DD}{{\mathcal D}}
\newcommand{\Sch}{{\mathscr S}}
\newcommand{\sD}{{\mathscr D}}

\newcommand{\HH}{{\mathcal H}}\newcommand{\LL}{{\mathcal L}}
\newcommand{\Hf}{H}
\newcommand{\CC}{\bC}
\newcommand{\aaa}{\tilde{\alpha}}

\newcommand{\ip}[2]{\langle #1 , #2\rangle}
\newcommand{\WF}{{\rm WF}}
\newcommand{\Had}{{\sf Had}}
\newcommand{\om}{\omega}

\renewcommand{\Re}{{{\rm Re}\,}}
\renewcommand{\Im}{{{\rm Im}\,}}

\newcommand{\II}{{\boldsymbol{1}}}

\begin{document}
\renewcommand{\thefootnote}{\fnsymbol{footnote}}
${}$ \par ${}$ \vspace{-2.5cm} ${}$ \par
\noindent
\begin{center}
{ \Large \bf Stability of Quantum Systems at Three Scales: \\Passivity, Quantum Weak Energy
  Inequalities\\[6pt] and the Microlocal Spectrum Condition}
\\[20pt]
{\large \sc Christopher J.\ Fewster${}^1$\ \ 
 {\rm and}\ \   Rainer Verch${}^{2,}$\footnote{Current address:
Max-Planck-Institute for Mathematics in the Sciences, Inselstr. 22,
04103 Leipzig, Germany. Email: verch@mis.mpg.de}}
\\[15pt]  
                 ${}^1$ Department of Mathematics,\\
                 University of York,\\
                 Heslington,\\
                 York YO10 5DD, United Kingdom\\[4pt]
                 e-mail: cjf3$@$york.ac.uk
                 \\[10pt]
                 ${}^2$\,Institut f\"ur Theoretische Physik,\\
                 Universit\"at G\"ottingen,\\
                 Bunsenstr.\ 9,\\
                 D-37073 G\"ottingen, Germany\\[4pt]
revised -- 3 April 2003
\end{center}
${}$\\[9pt]
{\small {\bf Abstract. }     
 Quantum weak energy inequalities have recently been
 extensively discussed as a condition on the dynamical stability of
 quantum field states, particularly on curved spacetimes. We formulate
 the notion of a quantum weak energy inequality for general dynamical
 systems on static background spacetimes and establish a connection
 between quantum weak energy inequalities and thermodynamics.
 Namely, for such a dynamical system, we show that the existence
 of a class of states satisfying a quantum weak inequality
 implies that passive states (e.g., mixtures of ground- and thermal equilibrium states) 
 exist for the time-evolution of the system and, therefore, that the 
 second law of thermodynamics holds. 
 As a model system, we consider the free scalar quantum field on a
 static spacetime. Although the Weyl algebra does not satisfy our
 general assumptions, our abstract results do apply to a related algebra
 which we construct, following a general method
 which we carefully describe, in 
 Hilbert-space representations induced by quasifree Hadamard states. 
 We discuss the problem of
 reconstructing states on the Weyl algebra from states on the new
 algebra and give conditions under which this may be accomplished.

 Previous
 results for linear quantum fields show that, on one hand, quantum weak energy
 inequalities follow from the Hadamard condition (or microlocal
 spectrum condition) imposed on the states, and on the other hand, that the
 existence of passive states implies that there is a class of states
 fulfilling the microlocal spectrum condition. Thus, the results of this paper
 indicate that these three conditions of dynamical stability are
 essentially equivalent. This observation is significant because the
 three conditions become effective at different length scales: The
 microlocal spectrum condition constrains the short-distance
 behaviour of quantum states (microscopic stability), quantum weak
 energy inequalities impose conditions at finite distance (mesoscopic
 stability), and the existence of passive states is a statement on the
 global thermodynamic stability of the system (macroscopic stability).
 }
${}$\newpage \noindent
\setcounter{footnote}{0}
\renewcommand{\thefootnote}{\arabic{footnote}}
\section{Introduction}
At small scales and high energies, the behaviour of matter is governed
by quantum field theory, and it is widely known that the
energy-momentum tensor in quantum field theory violates the
energy-positivity conditions that are generically fulfilled for
classical matter in 
general relativity. For instance, in flat spacetime, there is in any
quantum field theory a large class of physical states for which the
expectation value of the energy density is a function on spacetime
assuming negative as well as positive values \cite{EGJ}. However, as
pointed out by Ford \cite{FordNegEn}, one may argue on heuristic grounds
that there must be constraints on the intensity and spatio-temporal
extension of negative values of the energy density, as otherwise one
could produce situations in which macroscopic violations of the second
law of thermodynamics occur. This idea was further developed by Ford
and several other authors (see, e.g.,
\cite{FordRomanQInEq,FordRomanRest,PfenFord,AGWQI,EvesonFewster,FewsterTeo1,FewsterTeo2,Flan2d,Voll2,FV,PfEM})
 and led to a form
of such constraints which are now called {\it quantum weak energy
  inequalities} (abbreviated, QWEIs) in the terminology of \cite{FV},
or often simply {\it quantum inequalities} (QIs). To explain their
nature, consider some quantum field propagating on a Lorentzian
spacetime $(M,\gb)$, and let $\langle T_{\mu\nu}(x)\rangle_{\omega}$ be
the expectation value of the energy-momentum tensor in a state
(expectation functional)
$\omega$ at some spacetime point $x$. For any smooth, timelike curve
$\g$ in spacetime, denote by 
\[
 \rho_{\omega}(\tau) = \dot{\g}^\mu(\t) \dot{\g}^\nu(\t) \langle
 T_{\mu \nu}(\gamma(\t))\rangle_{\omega} 
\]
the mean energy density in the state $\omega$ along the curve,
parametrized by proper time $\t$. Then one says that the states of the
quantum field fulfill a QWEI if for any timelike  curve $\g$ and any
real-valued smooth, compactly supported test-function $g$ there holds
a bound of the form
\begin{equation} \label{qwei1}
\inf_{\omega}\, \int d\t \, g^2(\t)\r_{\omega}(\t) \ge c_{g,\g} > -\infty
\end{equation}
where the infimum is taken over the set of states $\o$ of the quantum
field theory in which $\r_{\o}$ can be reasonably defined as a
locally integrable function (typically, this is a dense set of all
physical states for the quantum field theory). The important point is that the
constant $c_{g,\g}$ bounding the weighted integral of the energy
density along $\g$ from below may depend on the weight-function $g$
and the curve $\g$, but is state-independent. 

To comment on this constraint, we note first that an inequality of the
form \eqref{qwei1} has been shown to hold for the class of Hadamard
states $\o$ of the free Klein-Gordon field and the Dirac field on
arbitrary, globally hyperbolic spacetimes \cite{AGWQI,FV}. Moreover,
in more special situations, the inequality \eqref{qwei1} was obtained
in a more specific form. For example, for the massless free scalar
field in $d$-dimensional Minkowski spacetime and the
specific choice of a (not compactly supported) Lorentzian weight
function
$g^2(\t) = t_0/[\pi(\t^2 + t_0^2)]$ $(t_0 > 0)$, and taking for $\g$ any
straight timelike line, one obtains a bound
\begin{equation} \label{qwei2}
\frac{t_0}{\pi}\int d\t\,\frac{\r_{\omega}(\t)}{\t^2 + t_0^2} \ge -
\frac{\G_d}{t_0^d}
\end{equation}
for all states $\omega$ of finite particle number and energy, where
$\G_d$  is a state-independent universal constant depending only on the
spacetime dimension \cite{FordRomanRest,Flan2d,EvesonFewster}.
Thus the intensity of the weighted
negative energy is at most proportional to an inverse power of its 
mean duration, i.e.\ the width $t_0$ of the Lorentzian weight
function. This considerably limits the possibility of negative
energies to build up to macroscopic violations of the second law of
thermodynamics. Also, other macroscopic dynamical instabilities are
hampered in this way when taking expectation values of a quantum
field's energy-momentum tensor as the right hand side of Einstein's
equations of gravity: It can be shown that appropriate forms of
averaged energy conditions such as~\eqref{qwei1} or \eqref{qwei2} imply similar
statements about the dynamical behaviour of solutions to
 Einstein's equations as do
the pointwise energy positivity conditions for classical matter.
Examples include 
singularity theorems, or the impossiblity of exotic spacetimes with
closed timelike curves or ``warp-drive'' scenarios
\cite{BarVis,FR-worm,PfenFordWarp}. 

The argument sketched above---that a bound on negative energy densities
in quantum field theory of the type~\eqref{qwei2} really prevents a
macroscopic violation of the second law of thermodynamics---is, of
course, somewhat heuristic. The purpose of the present article
is to show that such a conclusion may be drawn rigorously under fairly
general circumstances. To this end, we shall specialize our setting a
bit and consider a quantum system (which may, but
need not, be a quantum field) situated in a static, spatially
compact spacetime (the assumption of spatial compactness is mainly
made for convenience). Thus, the underlying spacetime $(M,\gb)$ is of the form
\[
M = \bR \times \Sigma\,,
\]
where $\Sigma$ is a smooth, $s$-dimensional compact manifold endowed
with a Riemannian metric $\hb$, and the static Lorentzian metric $\gb$
on $M$ has the line element
\begin{equation} \label{staticmetric}
 ds^2 = g_{ab}dx^adx^b = g_{00}dt^2 - h_{ij}dx^idx^j
\end{equation}
with a smooth, strictly positive function $g_{00}$ on
$\Sigma$. Consequently, when a point $x \in M$ is represented as $x =
(t,\ux)$ with $t \in \bR$ and $\ux \in \Sigma$, then $\gb$ is
independent of $t$. The variable $t$ is  to be viewed as
``time''-variable, whereas $\Sigma$   contains all the spatial
locations of the system at instances of ``equal time''.

We will then assume that the quantum system on this spacetime can be
modelled  such that the observables of the system are elements of a
$C^*$-algebra\footnote{We shall understand by a $C^*$-algebra
  always a $C^*$-algebra containing a unit element, generically
  denoted by {\bf 1}.} $\cA$, and that the time-evolution of the system can be
described by a one-parameter group $\{\a_t\}_{t\in\bR}$ of
automorphisms of $\cA$. A technical assumption is also made, namely
that the automorphism group is strongly continuous in time: This means
that $||\a_t(A) - A|| \to 0$ as $t \to 0$ for each $A \in \cA$; the norm
appearing here is the $C^*$-norm on $\cA$. In technical terms (cf.\
\cite{BR1}), the pair $(\cA,\{\a_t\}_{t\in\bR})$ is a $C^*$-dynamical
system.  Here the
time-parameter $t$ is thought of as having the significance of the
time-parameter of the spacetime, so that $\{\a_t\}_{t\in\bR}$
describes the time-evolution of the system on the underlying
spacetime. That is, if the system is formed by a family $\cA(O)
\subset \cA$, $O \subset \bR \times \Sigma$, of sub-$C^*$-algebras of
$\cA$ fulfilling the isotony condition $O_1 \subset O \imply \cA(O_1)
\subset \cA(O)$, then we assume that $\a_t(\cA(O)) = \cA(\t_t(O))$
where $\t_t$ denotes the time-shift $\t_t(t',\ux) = (t + t',\ux)$ on
the underlying spacetime.
These assumptions are very general (cf.\ \cite{Haag}); apart
from some points of technical detail (such as replacing the $C^*$-algebra
by a more general types of $*$-algebra, or relaxing the assumption of
strong continuity of the dynamics), there is a vast range of quantum
systems on static spacetimes that can be modelled as $C^*$-dynamical systems. 

Some comments on the extent to which the assumption of strong
countinuity is realized in quantum field theoretic system may
nevertheless be helpful. The typical description of such
systems in the present setting of a static spacetime $M$ starts
with a $C^*$-algebra $\fA$ generated by elements ${\sf G}(f)$, where
$f$ ranges over $\Coin(M)$. These algebraic generators are subject to
certain relations characterizing dynamical properties of the system,
and a time evolution can usually be defined on $\fA$ as a
one-parametric group $\{\aaa_t\}_{t\in\bR}$ of
automorphisms of $\fA$ via $\aaa_t({\sf G}(f))
= {\sf G}(f \circ \tau_{-t})$. However, thinking in particular of the
case of a free bosonic field, where ${\sf G}(f)$ are
the Weyl-generators of a CCR-algebra\footnote{See Sec.\ 4.1
  for a fuller discussion of the CCR-algebra in the context of the
  free Klein-Gordon field.}, the corresponding
$\{\aaa_t\}_{t\in\bR}$ will, in general, fail
to be strongly continuous. This difficulty can be overcome by passing
to sufficiently ``regular'' Hilbert-space representations $\pi$ of
$\fA$ in which $\{\aaa_t\}_{t\in \bR}$ is
weakly continuous. We will give a very detailed account of this
construction for the example of the scalar Klein-Gordon field in
Sec.~\ref{sect:qfsb}; to prepare the ground for this, we now outline the basic idea.

To simplify the discussion a little bit, and to help the reader who is
not too familiar with $C^*$-dynamical systems to see the connection
with the more common Hilbert space approach, we assume that there is a
weakly continuous unitary group $\{V_t\}_{t\in\bR}$ on $\cH$, the
representation Hilbert space of $\pi$, so that
$$ V_t \pi({\sf A}) V_t^* = \pi(\aaa_t({\sf A}))
$$
for all ${\sf A} \in \fA$. Then $\alpha_t(A) = V_t A V_t^*$, $A \in
\pi(\fA)$, defines a group of automorphisms of $\pi(\fA) \subset
B(\cH)$ which need not be strongly strongly continuous but only weakly
continuous, i.e.\
 $\langle \psi,(\a_t(B) - B)\phi\rangle \to 0$ as $t\to
 0$ for all choices of vectors $\phi,\psi \in \cH$. 
But this form of continuity implies that for each $h \in \Coin(\bR)$ and
each $A \in \pi(\fA)$ the weak (or Bochner-) integral
\begin{equation} \label{convdef}
 \alpha_hA = \int dt\,h(t)\alpha_t(A)
\end{equation}
exists as an element in $B(\cH)$; more precisely, $\alpha_hA$ is
contained in $\pi(\fA)''$, the von Neumann algebra generated by
$\pi(\fA)$.\footnote{For a subset $\cB$ of $B(\cH)$, $\cB'$ denotes the
  commutant of $\cB$, i.e.\ $\cB'=\{C \in B(\cH): CB = BC\ \forall\ B
  \in \cB \}$.} 
Moreover, it is easy to see (cf.\ Sec.~\ref{sect:qfsb}) that 
$|| \alpha_t(\alpha_hA) - \alpha_hA || \to 0$ for $t \to 0$. Hence, if we
denote by $\cA$ the $C^*$-algebra which is generated by all the
$\alpha_hA$ where $A \in \pi(\fA)$ and $h \in \Coin(\bR)$, then
$(\cA,\{\alpha_t\}_{t\in\bR})$ forms a $C^*$-dynamical system. (The
local algebras $\cA(O)$ are then defined as $\cA \cap \{\pi({\sf
  G}(f)): f \in \Coin(O)\}''$.)

Therefore one sees that $C^*$-dynamical systems arise naturally as a
means of describing quantum field systems as soon as
 one considers representations in
which the time-evolution of the system is modelled by a weakly
continuous unitary group (and also under more general circumstances,
see Sec.~\ref{sect:qfsb}), and this is surely a very general
situation, irrespective of whether a quantum field is interacting or
free. We should also emphasize that, once such a representation $\pi$
is chosen, it doesn't matter if one describes the system in terms of
the algebra $\pi(\fA)$ or the algebra $\cA$ since, for each unit
vector $\psi \in \cH$, the expectation value $\langle \psi,A\psi
\rangle$ of $A \in \pi(\fA)$ can be approximated as closely as desired
by the expectation value $\langle \psi,\alpha_hA \psi\rangle$ of
$\a_hA \in \cA$ through sharply peaking $h$ around $0$. Conversely,
$\a_hA$ is weaky approximated by elements of $\pi(\fA)$. For further
discussion of the relation between states on $\cA$ and states on $\fA$
in the case of the CCR-algebra, see towards the end of
Sec.~\ref{sect:dynsys} and
Appendix \ref{sect:corell}.

To make contact with thermodynamics, we now turn to the notion of
{\em passivity} introduced by Pusz and Woronowicz~\cite{PW}. Let
$(\cA,\{\a_t\}_{t\in\bR})$ be a $C^*$-dynamical system. The idea of
Pusz and Woronowicz was to consider the behaviour of the system when
the external conditions change in time. To this end, we note first
(cf.\ Sec.~\ref{sect:absres}) that
there exists a norm-dense $*$-subalgebra $D(\d)$ of $\cA$ so that 
\begin{equation} \label{derivation}
\d(A) = \left. \frac{d}{dt} \a_t(A)\right|_{t=0}\,, \quad A
\in D(\d)\,,
\end{equation}
exists in $\cA$. This derivation is the generator of the
dynamics when the external conditions remain unchanged. A change in
the external conditions can be modelled via replacing $\d$ by a
time-dependent dynamics-generator
\[
 \d_t(A) = \d(A) + i[H_t,A]\,, \quad A \in D(\d)\,,
\]
where $H_t = H_t^*$ is a smooth function of $t \in \bR$ having values in
$\cA$. Now assume that $H_t = 0$ for $t< 0$ and for $t > T$ where $T$
is some positive number. Then the dynamics of the system remains
unchanged before $t = 0$ and after $t=T$. In other words, the system
undergoes a cyclic change of external conditions during which it is
thermally isolated. (For $t<0$ and $t > T$, the system is closed.) It
can be shown that there is a unique smooth family $U_t^{\Hf}$, $t\in\bR$, of
unitary elements in $\cA$ solving the initial value problem 
\begin{equation}
\label{eq:QIVP}
 \frac{d}{dt}U_t^{\Hf} = \frac{1}{i}\a_t(H_t)U^{\Hf}_t\,,\ \ \ U^{\Hf}_0 = 1\,,
\end{equation}
and that, consequently, the family $\a^{\Hf}_t$, $t\in\bR$, of
automorphisms of $\cA$ given by $\a_t^{\Hf}(A) = U_t^{\Hf}{}^*\a_t(A)U_t^{\Hf}$
solves the initial value problem
\[
\frac{d}{dt}\a_t^{\Hf}(A) = \a_t^{\Hf}(\d_t(A))\,, \ \ \ \a_0^{\Hf}(A) = A\,,
\]
for all $A \in D(\d)$. 

We recall that a  state on a $C^*$-algebra $\cA$ is a continuous linear
functional $\o:\cA \to \bC$ which is positive, i.e.\ $\o(A^*A) \ge 0$
for all $A \in \cA$, and fulfills $\o({\bf 1}) = 1$. 

If the system is initially in the state $\o$,
then the work done on the system under the cyclic change of external
conditions is 
\[
 L^{\Hf}(\o) = \int_0^T dt\, \o(\a^{\Hf}_t(\frac{dH_t}{dt}))\,,
\]
and, as shown in~\cite{PW}, it holds that 
\begin{equation} \label{workform}
 L^{\Hf}(\o) = \frac{1}{i}\o(U^{\Hf}_T{}^*\d(U^{\Hf}_T))\,.
\end{equation}
Pusz and Woronowicz call a state $\o$ on $\cA$ {\it passive} if
$L^{\Hf}(\o) \ge 0$ for all smooth $H: \bR \owns t \mapsto H_t = H_t^* \in
\cA$ that have $H_t = 0$ for $t$ outside $[0,T]$ with some $T >
0$. Then it follows from \eqref{workform} that $\o$ is passive if and
only if
\[
     \inf_{U \in \cU_0(\d)}\,\frac{1}{i}\o(U^*\d(U)) \ge 0
\]
where $\cU_0(\d)$ denotes the set of all unitary $U$ in $D(\d)$ which
are in $\cA$ continuously connected to ${\bf 1}$.

Passive states may thus be viewed as states of the system which are
``in equilibrium'' in the sense that, if the system is in such a
state, then it is impossible to extract energy (gain work) from the
system under cyclic changes of the external conditions of the system
(while keeping it thermally isolated). It is in this sense that the
second law of thermodynamics is valid for passive states. However, the
set of passive states is really larger than the class of thermal
equilibrium states, as those have a definite temperature: Each mixture
of thermal equilibrium states (KMS-states) at arbitrary temperatures
(including zero, corresponding to ground states) is a
passive state. On the other hand, passive states are invariant under
$\{\a_t\}_{t\in\bR}$, and some (in most circumstances, mild)
additional conditions such as clustering, or complete passivity, imply
that passive states are, in fact, thermal equilibrium states at a
definite temperature, i.e.\ KMS-states or ground states. We refer to
\cite{PW} for a detailed discussion of these matters.
We should like to mention that, in the context of quantum field theory
in Minkowski-spacetime, Kuckert \cite{Kuckert1} has recently introduced
a weaker notion of passivity, called semipassivity, and has shown
that for stationary and homogeneous states semipassivity is equivalent
to the KMS-condition for a particular inertial observer. There are
also more recently investigated interconnections between energy-compactness,
semi-passivity, and thermal equilibrium properties of states in
quantum field theory \cite{Kuckert2,GL,BrosBu} to which we would like
to direct the reader's attention.

In the present work, we shall show that the existence of passive
states for $C^*$-dynamical systems modelled on a static,
spatially compact spacetime can be deduced from a suitable form of a
QWEI. For this purpose, we suppose that there exists an energy density
whose spatial integral yields the generator of the dynamics, and we
assume QWEIs for this energy-density to hold for a suitable dense set
of states on $\cA$. The precise assumptions will be 
discussed in Sec.~\ref{sect:absres}.
Given these assumptions, along with some other physically
well-motivated conditions, we obtain in Sec.~\ref{sect:absres} several assertions about the
existence of passive states for the $C^*$-dynamical system.
While the investigations in Sec.~\ref{sect:absres} are based on general,
model-independent assumptions,  Sec.~\ref{sect:qfsb} concerns the example
of the free scalar Klein-Gordon field on a static, spatially compact
spacetime. In particular, we will show that in the GNS-representation
of each quasifree Hadamard state one obtains a $C^*$-dynamical system
and an energy density on a suitable domain such that all the general
assumptions of Sec.~\ref{sect:absres} are satisfied. It is of course well-known that
quasifree Hadamard ground- and KMS-states exist for the free scalar field on static, spatially
compact spacetimes~\cite{Weinless}, but we emphasise that our arguments
do not make use of this fact. Thus 
there is the prospect that the results of Sec.~\ref{sect:absres} have a wider
applicability beyond the realm of free quantum fields. 
Instead, our arguments rely heavily on the
characterization of the Hadamard property in terms of a condition on
the wavefront set of the two-point function of a state for the free field,
called microlocal spectrum condition,
which was established in \cite{Rad1} (cf.\ also \cite{BFK,SV2}).
 However, we shall work with a
new version of that characterization taken in part from \cite{SVW} and
related to the concept of ``domain of microlocal smoothness'' in
\cite{BF}. Thus, Sec.~\ref{sect:qfsb} contains several apparently new microlocal
techniques in quantum field theory on curved spacetime. Some background
material on microlocal analysis, including a convenient calculus for
distributions taking values in Banach and Hilbert spaces is presented in
Sec.~\ref{sect:uloc}. 

To conclude this introduction, we would like to point out that the
results obtained in this work, together with results previously
obtained, indicate an intimate connection --- and apart from points of
technical detail, equivalence --- between the following classes of
states of a free quantum field (on static spacetimes):
\begin{itemize}
\item[(i)] states fulfilling the microlocal spectrum condition
  (Hadamard states),
\item[(ii)] states fulfilling QWEIs,
\item[(iii)] states induced by operations on passive states
  (corresponding to local, finite energy excitations of passive
  states)\,.
\end{itemize}
Namely, it was shown in \cite{AGWQI} (see also \cite{FV}) that free
quantum field states fulfilling the microlocal spectrum condition
satify QWEIs. In the present work, we show that
QWEIs imply the existence of passive states. On the other hand,
\cite{SV1} indicates that passive states fulfill the microlocal
spectrum condition (and this result carries over to a suitable dense
set of states in the GNS-representation of passive states, cf.\
\cite{Ver.acs}).

This observation is of considerable significance
since the regularity properties on the classes of states (i), (ii) and
(iii) are imposed on very different length scales. The states (i)
fulfill a constraint on their microscopic short-distance behaviour,
the condition on the states (ii) is imposed at finite length scales
while the specification of the class of states (iii) is a global
condition on their thermodynamic stability. We conjecture that this
connection between dynamical stability conditions at different length
scales is a generic feature and extends to physical systems described
by more general types of quantum fields.  

\section{Passivity from quantum weak energy inequalities}
\label{sect:absres}
We will consider a quantum system situated in a static spacetime of
the form $\bR \times \Sigma$, where $\Sigma$ is an $s$-dimensional
compact manifold; the variable $t \in \bR$ is interpreted as
``time''-variable, whereas $\Sigma$ contains all spatial locations of
the system. 
Furthermore, we assume that the spacetime is endowed with a static
metric of the form \eqref{staticmetric}. Then this metric induces a
preferred measure on the
Borel-sets of $\Sigma$ which we will denote by 
$d\mu$; it is the volume measure of the Riemannian metric $\boldsymbol{h}$ on
$\Sigma$. In local coordinates $(\ux^i)$ for $\Sigma$, $d\mu(\ux)$ has the
coordinate expression $\sqrt{{\rm det}(h_{ij})}\, d^s\ux$.
(We note that everywhere in the discussion to follow one could replace
$d\mu$ by any other Borel-measure on $\Sigma$, but the choice just
made is convenient and natural.) 

We shall make some general, model-independent assumptions concerning
the mathematical description of the dynamical properties of the
system, suited to discuss the connection between QWEIs and
passivity. In Section~\ref{sect:qfsb} we will show that all these
assumptions are 
fulfilled in the case of the free scalar Klein-Gordon field on a
static, globally hyperbolic spacetime. 
\\[6pt]
Our quantum system will be modelled by a $C^*$-dynamical system 
$(\cA,\{\a_t\}_{t\in\bR})$ as discussed in the Introduction.
\\[6pt]
Since $\{\a_t\}_{t\in\bR}$ is strongly continuous, there exists a
norm-dense (maximal) domain $D(\d)$ in $\cA$ so that the derivation
$$ \d(A) = \left.\frac{d}{dt}\a_t(A)\right|_{t= 0} $$
exists in $\cA$ for all $A \in D(\d)$. In fact, $D(\d)$ is a norm-dense
$*$-subalgebra of $\cA$, as is the set $D^{\infty}(\d)$ of all
elements in $\cA$ such that $t \mapsto \a_t(A)$ is $C^{\infty}$ at
$t=0$ (i.e., the common domain for $\d^n$, for all $n \in \bN$), 
since defining for each $A \in \cA$ and $h \in
C^{\infty}_0(\bR)$ their convolution $\a_hA$ with respect to
$\{\a_t\}_{t\in\bR}$ by \eqref{convdef}
(with the integral now converging in norm because of strong continuity)
one finds that $\a_hA \in D^{\infty}(\d)$.

Next, we need to introduce an energy-density in order to connect our
dynamical system with QWEIs. Since energy-densities are in many
examples unbounded operators or quadratic forms, we need an
appropriate domain for such a quantity with convenient algebraic and
density properties.
Therefore, we shall now make the assumption that there is a 
norm-dense subspace $\cW^{\infty} \subset
D^{\infty}(\d)$ which has the property of being stable under
taking adjoints and under
convolution with test-functions with respect to the action of
$\{\a_t\}_{t\in\bR}$. In other words, for all $A \in \cW^{\infty}$ and
$f \in C^{\infty}_0(\bR)$, $\a_fA$ is also contained in
$\cW^{\infty}$.
We shall denote by $\cA^\infty$ the $*$-algebra generated by $\cW^\infty$.

It will turn out to be convenient to introduce
the set $\cU^{\infty} \subset D^{\infty}(\d)$ of all unitaries $U$ which
are of the form
\[
U = {\rm e}^{iA_1} \cdots {\rm e}^{iA_N} 
\]
where $N \in \bN$ and $A_1,\ldots,A_N$ are hermitean elements in
$\cA^\infty$. Lemma~\ref{L2} below will show that it
suffices to consider $U$ of this form when discussing general cyclic
changes to the system such as that described by 
$U_t^{\Hf}$ in Eqs.~\eqref{eq:QIVP} and~\eqref{workform}. 
 We denote by $\cU_{\rm alg}^{\infty}$ the
$*$-algebra which $\cU^{\infty}$ generates; this is a sub-$*$-algebra
of $D^{\infty}(\d)$. (In fact, as a consequence of Lemma 2.2 below, the
closures of $\cU^{\infty}_{\rm alg}$ and $D^{\infty}(\d)$ in the
graph-norm of $\d$ coincide.)
\\[6pt]
Now we can formulate our assumptions on the existence of an
energy density generating the dynamics: We
assume that there exists a set 
$\cS$ which is a subset of the set of states on $\cA$ and 
has the property of being
closed under finite convex combinations and operations induced by
elements of $\cU_{\rm alg}^{\infty}$, 
i.e.\ if $\varphi \in
\cS$ and $A \in \cU^{\infty}_{\rm alg}$,
 then the state $\varphi^A$ on $\cA$ defined by
\[ \varphi^A(B) := \varphi(A^*BA)/\varphi(A^*A)\,, \quad B \in
\cA\,,
\]
is also contained in $\cS$. We denote by $\cV$ the
vector-space generated by $\cS$ and $\cU^{\infty}_{\rm alg}$,
 that is, the subset of all elements
$\ell$ in the continuous dual space $\cA^*$ of $\cA$ that arise as
finite linear combinations of  elements of $\cS$ operated upon
by elements of $\cU^{\infty}_{\rm alg}$ from right and left: 
\begin{equation}
\label{eq:Vdef}
\ell(B) = \sum_{i=1}^n \varphi_i(A_iBC_i) \qquad n \in \bN,~ 
A_i,B_i \in \cU^{\infty}_{\rm alg}~{\rm and}~\varphi_i \in \cS\,.
\end{equation}

The {\it energy density} of the system is then
defined to be 
a linear map $\rrb$ taking elements in $\cV$ to $C^1$-functions on
$\bR \times \Sigma$. That is, given $\ell \in
\cV$, then $\rrb[\ell]$ is a $C^1$-function on $\bR \times \Sigma$,
 and the assignment $\ell
\mapsto \rrb[\ell]$ is linear. We will find it convenient to
write $\ell(\rrb(t,\ux))$ for $\rrb[\ell](t,\ux)$. Moreover, we will
write $\ell([\rrb(t,\ux),A])$ to denote $\rrb[\ell_A -
{}_A\ell](t,\ux)$, where the functionals $\ell_A$ and ${}_A\ell$ are
given by
\[
\ell_A(B) = \ell(BA)\,, \quad {}_A\ell(B) = \ell(AB)\,.
\]
The quantity $\varphi(\rrb(t,\ux))$ 
ought to be viewed as the expectation value of the
energy-density with respect to the preferred time-coordinate in the
state $\varphi \in \cS$ at $(t,\ux)$.
The following assumptions are in line with this point of view:
\begin{itemize}   
\item[(i)] It will be assumed that
 $$ \int_{\Sigma} d\m(\ux) \ell([\rrb(t,\ux),\a_t(A)]) =
 -i\frac{d}{dt}\ell(\a_t(A))\,, \quad A \in \cU^{\infty}_{\rm alg}
\,,\ \ell \in \cV\,.$$
\item[(ii)] It will also be assumed that the integrated generating
  term of the energy density
  is conserved:
    $$ \frac{d}{dt}\int_{\Sigma} d\m(\ux)\,\ell([\rrb(t,\ux),A]) = 0\,,
    \quad A \in \cU^{\infty}_{\rm alg}\,,\ \ell \in \cV\,.$$
\end{itemize}
We note that, in the presence of (i), condition (ii) may be equivalently
expressed by saying that the generators of the time-evolutions are
independent of $t$,
$$ \frac{d}{dt}\ell(\a_t(A)) = \ell(\d(\a_t(A)))\,,
 \quad A \in \cU^{\infty}_{\rm alg}
\,,\ \ell \in \cV\,.$$
${}$\\
With the assumptions formulated so far, the notion of quantum weak
energy inequality in the present general setting can now be precisely defined.
\begin{Definition} \label{D1}
Assume that a physical system on the static spacetime $\bR \times
\Sigma$ modelled by $\cA$, $\{\a_t\}_{t\in\bR}$, $\cW^\infty$, $\cA^{\infty}$,
$\cS$ and $\rrb$ with the properties stated above is given, and let
$\omega \in \cS$.
\begin{itemize}
\item[(a)] We say that $\o$ fulfills a {\bf static quantum weak energy
  inequality} with respect to $\cS$ if there is for each real-valued $g
  \in C_0^{\infty}(\bR)$ a locally integrable\footnote{That is, each
$\ux\in\Sigma$ should have an open neighbourhood on which $q(g;\cdot)$
is integrable.}
 non-negative function $\Sigma \owns \ux \mapsto
  q(g;\ux)$ such that 
\begin{equation}
\label{qWEI}
\int_{\bR} dt\, g^2(t)\,\varphi(\rrb(t,\ux)) - \int_{\bR}dt \,
g^2(t)\o(\rrb(t,\ux)) \ge - q(g;\ux)
\end{equation}
holds for all states 
$\varphi \in \cS$ and all $\ux \in \Sigma$.
\item[(b)] We say that $\o$ fulfills a {\bf limiting static QWEI} (with
  respect to $\cS$) if
  $\o$ fulfills the static QWEI \eqref{qWEI}, and in addition each
$\ux\in\Sigma$ has an open neighbourhood $U$ such that
\begin{equation}
\Gamma_{U}:= \sup_g \limsup_{\lambda\to 0^+}
\frac{1}{\|g_\lambda^2\|_{L^1}}
\int_U d\mu(\ux')\, q(g_{\lambda};\ux')
<\infty
\label{eq:LSQWEI}
\end{equation}
where the supremum is taken over real-valued $g\in\Coin(\RR)$ with $\|g^2\|_{L^1}\not=0$, 
and $g_{\lambda}(t) = g(\lambda t)$. 
\item[(c)] A state $\o \in \cS$ will be called {\bf quiescent} (with
  respect to $\cS$) if $\o$ fulfills a limiting static QWEI in which
each $\Gamma_{U}=0$. (Since $q(g,\ux)$ is non-negative this is in fact
equivalent to the assertion that each $\ux\in\Sigma$ has an open
neighbourhood $U$ such that 
 $$ \lim_{\lambda \to 0^+} \,\lambda \int_U d\mu(\ux')\,q(g_{\lambda};\ux')=0$$
for each real-valued $g \in C_0^{\infty}(\bR)$.)
\end{itemize}
\end{Definition}
{\it Remarks. } \\[2pt]
(i) It is easy to see that, if there exists a state $\o \in \cS$
fulfilling a static QWEI with respect to $\cS$, then all states $\o'
\in \cS$ satisfy a static QWEI with respect to $\cS$ as
well. Therefore, it is actually more appropriate to say that the set
of states $\cS$ fulfills a static QWEI. Within the class of states
$\cS$, the condition of static QWEI is thus independent of the
individually chosen state. In contrast, conditions {\it (b)} and {\it (c)} are
state-dependent.
\\[6pt]
(ii) The element $\ux \in \Sigma$ appearing in $q(g;\ux)$ ought to
viewed as a label for the timelike curve $\gamma_{\ux}(t) =
\t_t((0,\ux))$, i.e.\ the orbit of the point $(0,\ux) \in M$
(identified with $\ux \in \Sigma$) under the one-parametric group of
time-shifts. Thus $q(g;\ux)$ (or rather, $\tilde{q}(g;\ux) = q(g;\ux)
+ \int dt\,g^2(t)\o(\rrb(t,\ux))$) in \eqref{qWEI} plays exactly the
role of $c_{g,\gamma_{\ux}}$ in \eqref{qwei1}.
\\[6pt]
(iii) The condition {\it (a)} given above is a 
 purely local statement about the 
dynamical system, and {\it (b)} and {\it (c)} are limits thereof which
are still ``spatially local''. 
However, since $\Sigma$ is compact and $q(g;\cdot)$ is
non-negative, one may easily draw global consequences: if a static QWEI
holds, then $q(g;\cdot)\in L^1(\Sigma,d\mu)$, while if a limiting static QWEI
holds then~(\ref{eq:LSQWEI}) also holds for $U=\Sigma$,
with $\Gamma_\Sigma=0$ in the
quiescent case. (Conversely, these global properties of course imply the
local statements given above.) 
It is these global statements which will appear in the
arguments presented below. 
\\[6pt]
(iv) It will be shown in Section~\ref{sect:qfsb} that any 
state for the scalar Klein-Gordon field on a static, globally
hyperbolic spacetime possessing a two-point function that is
stationary and of Hadamard form fulfills a limiting static QWEI with
$$ 
q(g;\ux) = \int_{\bR} du\,|\widehat{g}(u)|^2Q(u,\ux) 
$$
where $\widehat{g}$ denotes the Fourier-transform of $g$, and $Q$ is a
non-negative measurable function on $\bR \times \Sigma$ so that
$Q(u,\ux)$ is polynomially bounded in $u$ for each fixed $\ux \in
\Sigma$. Moreover, $Q(u,\ux)$ is $d\mu$-integrable over $\Sigma$ with
respect to $\ux$ for each $u\in\bR$, and
$\int_{\Sigma}d\mu(\ux)\,Q(u,\ux)$ is polynomially bounded in
$u$. This then implies that all states of the scalar Klein-Gordon
field on a stationary, globally hyperbolic spacetime whose two-point
functions are of Hadamard form (or, synonymously, satisfy the
microlocal spectrum condition) fulfill a static QWEI.
\\[10pt]
We have now collected all the assumptions relevant for the present
section. Before presenting our results on passivity properties of
states satisfying static weak energy inequalities for systems obeying
the assumptions given above, we put on record an
auxiliary lemma. Recall (cf.\ Introduction) that 
$\cU_0(\d)$ denotes the set
of all unitaries in $D(\d)$ which are continuously connected to the
unit ${\bf 1}$, i.e.\ those unitaries $U$ in $D(\d)$ so that there
is a norm-continuous curve $[0,1] \owns t \mapsto U(t)$ of unitaries
in $\cA$ so that $U(0) = {\bf 1}$, $U(1) = U$.
\begin{Lemma} \label{L2}
$\cU^{\infty}$ is dense in $\cU_0(\d)$ with respect to the graph-norm
of $\d$; i.e.\ for each $U \in \cU_0(\d)$ there is a sequence $U_n \in
\cU^{\infty}$, $n\in\bN$, with
$$ ||\,U_n - U\,|| + ||\,\d(U_n) -\d(U)\,|| \to 0 \quad \text{as}\ n
\to \infty\,.$$
\end{Lemma}
{\it Proof. } Let $A$ be a hermitean element in $D(\d)$. We show
that there is a sequence of hermitean elements in $\cA^{\infty}$
approximating $A$ in the graph-norm of $\d$. To this end, pick
$\epsilon > 0$ arbitrarily. Then choose some real-valued, non-zero $f \in
C_0^{\infty}(\bR)$ so that
$$ || \a_fA - A || + || \a_f\d(A) - \d(A) || < \epsilon/2\,.$$
Because $\cW^\infty \subset\cA^{\infty}$ is norm-dense in $D(\delta)$
and stable under taking adjoints, one may also choose some
hermitean $A_{\epsilon} \in \cW^{\infty}$ so that
$$ || A_{\epsilon} - A || < \frac{\epsilon}{2}(||f||_{L^1} +
||\dot{f}||_{L^1})^{-1} $$
where $\dot{f}$ is the derivative of $f$. Then one estimates
\begin{eqnarray*}
\lefteqn{|| \a_fA_{\epsilon} - A|| + ||\d(a_fA_{\epsilon}) -\d(A)||}\\
 &\le& ||\a_fA_{\epsilon} -\a_fA || + ||\a_fA -A || +
 ||\d(\a_fA_{\epsilon}) - \d(\a_fA)|| + ||\a_f\d(A) - \d(A)|| \\
&\le& \epsilon/2 + ||f||_{L^1}||A_{\epsilon}-A|| +
||\dot{f}||_{L^1}||A_{\epsilon} -A|| \ \le \ \epsilon\,,
\end{eqnarray*}
by making use of $\a_f\d(B) = \d(\a_fB) = -\a_{\dot{f}}(B)$ for all $B
\in D(\d)$ and $||\a_gB||\le ||g||_{L^1}||B||$ for
$g\in C_0^{\infty}(\bR)$.
As $\cW^{\infty}$ is by assumption stable under convolution with
test-functions with respect to the dynamical automorphism group, this
shows that each hermitean $A \in D(\d)$ can be approximated by a
sequence of hermitean elements in $\cW^\infty \subset\cA^{\infty}$ 
in the graph-norm of $\d$.

As argued in the bottom part of page 279 in \cite{PW} (cf.\ also Thm.\
5.4.28 in \cite{BR2}), for each $U \in
\cU_0(\d)$ there exist finitely many hermitean elements
$A_1,\ldots,A_N \in D(\d)$ with $||A_j|| < \pi$ ($j=1,\ldots,N$) so
that 
$$ U = {\rm e}^{iA_1} \cdots {\rm e}^{iA_N}\,.$$
Since each $A_j$ may be approximated by a sequence of hermitean
elements $A_j^{(n)}$, $n\in \bN$, in $\cA^{\infty}$, it is quite easy
to see that also $U_n = {\rm e}^{iA^{(n)}_1} \cdots {\rm
  e}^{A^{(n)}_N}$ approximates $U$ in the graph-norm of $\d$, and each
$U_n$ is contained in $\cU^{\infty}$. This proves the lemma. ${}$
\hfill $\Box$
\\[6pt]
Now, with the notation and assumptions introduced prior to Lemma 2.2,
our first result reads as follows.
\begin{Theorem} \label{P3}
Suppose that a state  $\o \in \cS$ satisfies a static QWEI. Then 
there exists a state $\o^{p}$ on $\cA$ which is passive for the
automorphism group $\{\a_t\}_{t\in\bR}$, i.e.\ it fulfills
$$ \frac{1}{i}\o^{p}(U^*\d(U)) \ge 0$$
for all the unitaries $U$ in $\cU_0(\d)$.

Moreover, if $\o$ is a quiescent state, then it is a passive state for
the automorphism group $\{\a_t\}_{t\in\bR}$.
\end{Theorem}
The proof of the first part of that result is based on a simple
\begin{Lemma} \label{L4}
Let $\o$ be a state on $\cA$ so that
$$ c_{\o}:= \inf_{U \in \cU^{\infty}}\,\frac{1}{i}\o(U^*\d(U)) >
-\infty\,,$$
then there exists a state $\o^{p}$ on $\cA$ which is passive for
$\{\a_t\}_{t\in \bR}$.
\end{Lemma}
{\it Proof of Lemma \ref{L4} } If $c_{\o} \ge 0$, then $\o$ is already
passive. Therefore, we assume that $c_{\o} \in (-\infty,0)$. There
exists a sequence $\tU_n \in \cU^{\infty}$, $n \in \bN$, so that 
$\frac{1}{i}\lim_{n\to \infty}\o(\tU_n^*\d(\tU_n))
 = c_{\o}$. The sequence of states
 $\o_n$, $n\in \bN$, on $\cA$ given by $\o_n(A) = \o(\tU^*_nA\tU_n)$
 possesses, by the Banach-Alaoglu-theorem \cite{RS1}, weak-* limit
 points, i.e.\ there is a state $\o^p$ on $\cA$ and a subnet
 $\{\o_{n(\s)}\}_{\s \in S}$ of $\{\o_n\}_{n\in\bN}$ so that
$\lim_{\s} \o_{n(\s)}(A) = \o^p(A)$ for all $A \in \cA$. 
Abbreviating $\tU_{n(\s)}$ by $\tU_{\s}$, we obtain for all $U \in
\cU^{\infty}$
\begin{eqnarray*}
 \lefteqn{\frac{1}{i}\o^p(U^*\d(U)) = \frac{1}{i}
\lim_{\s}\,\o(\tU_{\s}^*U^*\d(U)\tU_{\s})}\\
 & = & \frac{1}{i}\lim_{\s}\,\left( \o((U \tU_{\s})^*\d(U\tU_{\s})) -
   \o(\tU_{\s}^*\d(\tU_{\s}))\right)\\
 & \ge & \liminf_{\s}\frac{1}{i}\o((U\tU_{\s})^*\d(U\tU_{\s})) -
 \frac{1}{i}\lim_{\s}\o(\tU^*_{\s}\d(\tU_{\s}))\\
& \ge & c_{\o} - c_{\o} = 0\,.
\end{eqnarray*}
In view of Lemma \ref{L2}, this relation entails that 
$$ \inf_{U \in \cU_0(\d)}\,\frac{1}{i}\o^p(U^*\d(U)) \ge 0\,,$$
showing that $\o^p$ is passive. {}
 \hfill $\Box$
\\[10pt]
{\it Proof of Thm.\ \ref{P3} } In view of conditions (i) and (ii) 
one obtains for all $U \in \cU^{\infty}$ and any real $g \in
C^{\infty}_0(\bR)$ with $||g^2||_{L^1} > 0$, 
\begin{eqnarray*}
\lefteqn{\frac{1}{i}\o(U^*\d(U)) = \int_{\Sigma}d\m(\ux)\,
\o(U^*[\rrb(t,\ux),U])}\\
& = & \frac{1}{||g^2||_{L^1}}\int_{\bR}dt\,
g^2(t)\int_{\Sigma}d\m(\ux)\,\left(\o(U^*\rrb(t,\ux)U) -
  \o(\rrb(t,\ux))\right) \\
& = & \frac{1}{||g^2||_{L^1}}\int_{\Sigma}d\m(\ux)\int_{\bR} dt\,
g^2(t)\left(\o(U^*\rrb(t,\ux)U) -
  \o(\rrb(t,\ux))\right) \\
& \ge &  \frac{-1}{||g^2||_{L^1}}\int_{\Sigma}d\m(\ux) q(g;\ux) \,.
\end{eqnarray*}
If $\o$ is quiescent, we replace $g$ by $g_\lambda$ in the last term
and take the limit as $\lambda\to 0^+$ in order to conclude that $\o$ is
passive by Lemma \ref{L2}. 
Otherwise, we nonetheless have $\inf_{U\in\cU^\infty} i^{-1}
\o(U^*\d(U)) >-\infty$; the existence of the passive state $\o^p$
follows from 
Lemma \ref{L4}. 
\hfill $\Box$
\\[10pt]
It is shown in \cite{PW} that if $\o$ is a passive state which is
weakly clustering in time (and non-central) then $\o$ is a ground
state or a KMS-state at positive inverse temperature for the
time-evolution $\{\a_t\}_{t\in\bR}$. Here we add another observation,
assuming that a state fulfilling a limiting static QWEI 
is weakly clustering in time for the
integrated energy density. We denote by
$$ \ell(\rrb_{\Sigma}(t)) = \int_{\Sigma}d\m(\ux)\,
\ell(\rrb(t,\ux))\,, \quad \ell \in \cV\,,$$
the {\it integrated energy density} at time $t$ in the functional
$\ell$, and we will say that a state $\o \in \cS$ is {\it weakly
  clustering in time for the integrated energy density} if for every
$B \in \cU^{\infty}_{\rm alg}$ there
exists a sequence $\{\lambda_n\}$ of positive real numbers converging
to $0$ and some real-valued $g \in C^{\infty}_0(\bR)$ with $\int
dt\,g_n^2(t) =1$  so that 
\begin{equation}
\label{cluster}
\int_{\bR}dt\, \lambda_n g^2(\lambda_nt)\{\o(B\rrb_{\Sigma}(t)) -
\o(B)\o(\rrb_{\Sigma}(t))\} \to 0\ \ {\rm as}\ \ n \to \infty\,.
\end{equation}
We need another piece of notation. Recall that any state $\o$ on a
$C^*$-algebra $\cA$ induces the so-called GNS-representation
$(\cH,\pi,\O)$ consisting of a Hilbert-space $\cH$, a linear
$*$-representation $\pi$ of $\cA$ by bounded linear operators on
$\cH$, and a unit vector $\O$ which is cyclic for $\pi(\cA)$ in the
sense that $\pi(\cA)\O$ is dense in $\cH$, and which fulfills $\o(A) =
\langle \O,\pi(A)\O\rangle$ for all $A \in \cA$. If in addition $\o$
is invariant with respect to the automorphism group
$\{\a_t\}_{t\in\bR}$, $\o \circ \a_t = \o$, then there is a strongly
continuous unitary group $V_t$, $t \in \bR$, on $\cH$ which implements
the action of $\alpha_t$, $t\in\bR$, in the GNS-representation:
$V_t\pi(A)V_t^* = \pi(\a_t(A))$, $A\in\cA$; moreover, $V_t\O = \O$ for
all $t$ (see, e.g.\ \cite{BR1} for a thorough discussion of these
matters).
 The selfadjoint operator $H$ in $\cH$ with $V_t = {\rm
  e}^{iHt}$ is called the generator of the implementing unitary group.
\begin{Theorem} \label{P5}
  Let $\o\in\cS$ be a state which is weakly clustering in time for
  the integrated energy density.
\begin{itemize}
\item[(a)] If $\o$ is a quiescent state, then it follows that $\o$ 
is a ground state for  $\{\a_t\}_{t\in\bR}$.
\item[(b)] If $\o$ is invariant with respect to the automorphism group
  $\{\a_t\}_{t\in\bR}$ and fulfills a limiting static QWEI, 
  then the generator $H$ of the implementing unitary group
  is bounded below by $-\Gamma_{\Sigma}$, defined by
Eq.~(\ref{eq:LSQWEI}) [see also remark~(iii) following
Def.~\ref{D1}]. 
\end{itemize}
\end{Theorem}
{\it Proof. } Choose any
$A\in\cU^{\infty}_{\rm alg}$ with $\o(A^*A)= 1$ and let
$\{\lambda_n\}$ be a sequence of positive real numbers converging to
$0$ as well as  $g \in C_0^{\infty}(\bR)$ with $||g^2||_{L^1} = 1$ such
that \eqref{cluster} holds with $B = A^*A$. Then, setting $g_n(t) =
\sqrt{\lambda_n} g(\lambda_n t)$ it follows that
$$ \epsilon_n := \int_{\bR}dt\,g_n^2(t)\{\o(\rrb_{\Sigma}(t)) -
\o(A^*A\rrb_{\Sigma}(t))\}\to 0 \ \ {\rm as}\ \ n\to\infty\,.$$
Recalling that $\o(A^*[\rrb_{\Sigma}(t),A])$ is independent of $t$,
there results the following chain of relations:
\begin{eqnarray*}
 \frac{1}{i}\o(A^*\d(A)) &=& 
 \o(A^*\rrb_{\Sigma}(t)A) - \o(A^*A\rrb_{\Sigma}(t))\\
& = &
\int_{\bR} dt\, g_n^2(t)\{\o(A^*\rrb_{\Sigma}(t)A) - 
\o(A^*A\rrb_{\Sigma}(t))\} \\
& = &
\int_{\bR}dt\,g_n^2(t)\{\o(A^*\rrb_{\Sigma}(t)A)
 - \o(\rrb_{\Sigma}(t))\} + \epsilon_n \\
&\ge &
 -  \int_{\Sigma} d\mu(\ux)\,
 \frac{q(g_{\lambda_n};\ux)}{||g_{\lambda_n}^2||_{L^1}} 
 +\epsilon_n\,.
\end{eqnarray*}
In the limit $n\to\infty$, this gives
\begin{equation}
\frac{1}{i}\o(A^*\d(A)) \ge -\limsup_{\lambda\to 0^+} 
\int_{\Sigma} d\mu(\ux)\,
\frac{q(g_\lambda;\ux)}{\|g_\lambda^2\|_{L^1}}\ge -\Gamma_\Sigma\,,
\label{eq:AsdAli}
\end{equation}
since in either case $\o$ obeys a limiting static QWEI.
\\[6pt]
Case $(a)$. As $\o$ is assumed to be quiescent, $\Gamma_\Sigma=0$, so 
$$ \frac{1}{i}\o(A^*\d(A)) \ge 0$$
holds for all $A \in \cU^{\infty}_{\rm alg}$ with $\o(A^*A)=1$,
hence also for all $A \in \cU^{\infty}_{\rm alg}$. The proof of Lemma
\ref{L2} shows that 
the closure of $\cU^{\infty}_{\rm alg}$ in the graph-norm of $\d$
contains $D(\d)$, and thus $\o$ is a ground state. 
\\[6pt]
Case $(b)$. In
view of the definition of $\Gamma_{\Sigma}$, Eq.~(\ref{eq:AsdAli}) entails
$$ \langle A\O,H A \O\rangle = \frac{1}{i}\o(A^*\d(A)) \ge -\G_{\Sigma} $$
for all $A \in \cU_{\rm alg}^{\infty}$ with $||A\O||^2=1$, 
and hence also for all $A \in
D(\d)$ with $||A\O||^2 =1$. Since $D(\d)\O$ contains a core for
$H$, it follows that $H$ is bounded below by $-\G_{\Sigma}$. {}
\hfill $\Box$ 
\\[10pt]
Thm.\ \ref{P3} shows that the existence of passive states for the
given system may be deduced from the existence of states satisfying a
static quantum weak energy inequality; however, this result leaves room for
further sharpening. In particular, one would like to know if the
passive state $\o^p$ is normal to the state $\o$ assumed to fulfill a
static quantum weak energy inequality. It turns out that such an
assertion can be made under the assumption that $\o$ satisfies a
certain condition of ``energy compactness''. Let us fix the required
assumptions in detail.

We assume that we are given a Hilbert-space $\cH$ and a strongly
continuous unitary group $\{V_t\}_{t\in\bR}$ on $\cH$ with selfadjoint
generator $H$, i.e.\ $V_t = {\rm e}^{itH}$. Furthermore, we assume
that there is a $C^*$-subalgebra $\cA$ of $\cL(\cH)$ which contains the
unit operator ${\bf 1}$ and is left invariant under the automorphism
group $\a_t := {\rm Ad}V_t$, $t\in\bR$. In addition, it will be
assumed that $\{\a_t\}_{t\in\bR}$ acts strongly continuously on the
elements of $\cA$, and that, for each $E\ge 0$, the spectral projector 
$P_E$ of $H$ corresponding to the
spectral interval $[-E,E]$, is also contained in $\cA$.
 With these conventions, the
definitions of $\d$, $D(\d)$, $\cU^{\infty}$ and $\cU^{\infty}_{\rm
  alg}$, etc., are as before.  A vector $\O \in \cH$ will be said to
be {\it energy-compact} if for each finite $E > 0$ the set
$P_E\cA_{(1)}\O$ is a pre-compact subset of $\cH$. Here $\cA_{(1)}$ is
the set of all elements in $\cA$ whose norm is bounded by 1, and we
recall that a subset $\cJ$ of $\cH$ is called pre-compact if each
sequence $\{\chi_n\}_{n\in\bN}$ in $\cJ$ possesses a
sub-sequence $\{\chi_{n(k)}\}_{k\in\bN}$ which converges strongly to
some element $\chi \in \cH$, i.e.\ $||\chi_{n(k)} -\chi|| \to 0$ as $k
\to \infty$.

Energy-compactness conditions were introduced by Haag and Swieca
in quantum field theory  \cite{HaagSwieca}; they impose restrictions on
the energy-level density of quantum states. The original approach of
Haag and Swieca has been considerably extended and refined to
so-called ``nuclearity conditions'' in quantum field theory
in a series of works by Buchholz and collaborators.
It is interesting to note that there is a close connection between
such nuclearity conditions and decent thermodynamical properties of
quantum field systems. In fact, this was one of the central
motivations for the introduction of nuclearity conditions in \cite{BuchWich}.
 We recommend
\cite{Haag} and \cite{BuchWich,BuPo,Sum.Yu,Sum.Rev}
 for further discussions and
references on that subject, and we would like to refer also to
\cite{BrosBu,GL,Kuckert1,Kuckert2} for some recent developments.
Here, we just mention that related
energy-compactness conditions have been shown to hold for energy-ground
states (vacuum states) in several quantum
field theoretical models (see the quoted works, and references quoted
therein); notably, they have also been established for 
linear quantum field theories on ultrastatic curved spacetimes
\cite{HolDan.DiracNuc,Ver.KGNuc}.

The assumptions listed above now lead to
\begin{Theorem} \label{P6}
Let $\O$ be a unit vector in $\cH$ which is energy-compact, and let
$\o(A) := \langle\O,A\O\rangle$ be the vector-state on $\cA$ induced
by $\O$. If  
$$ c_{\o} := \inf_{U \in \cU_0(\delta)}\,\frac{1}{i}\o(U^*\d(U)) >
-\infty\,,$$
then there exists a unit vector $\O^p \in \overline{\cA\O}$ so that the
vector-state $\o^{p}$ induced by $\O^p$ is passive.
\end{Theorem}
{\it Proof. } There is a sequence $\tU_n$, $n\in\bN$, in $\cU^{\infty}$
so that $\lim_{n\to \infty}\frac{1}{i}\o(\tU^*_n\d(\tU_n)) =
c_{\o}$. Since $\tU_n\O$, $n\in\bN$, is a bounded sequence in $\cH$ and
possesses a weakly converging sub-sequence, it is no loss of
generality to assume that $w$-$\lim_{n\to\infty}\,\tU_n\O = \O^p$ for some
unit vector $\O^p \in\cH$.
 Now let $A_1,\ldots,A_N$ be finitely many 
hermitean elements in $D^{\infty}(\d)$, define $U$ as in the proof of
Lemma~\ref{L2}, 
 and write
$$ U_E := {\rm e}^{iP_EA_1P_E}\cdots {\rm e}^{iP_EA_NP_E}\,, \quad E > 0\,.$$
Then it holds that $U_E = {\bf 1} + G_E$ where $||G_E|| \le 2$ and 
$$ G_E = P_EG_EP_E\,.$$
Introducing the bounded operator $F_E:= ({\bf 1} + G_E^*)[H,G_E] =-i
U_E^*\d(U_E)$, we note that $F_E = P_EF_EP_E$. Moreover, by assumption
$P_E\cA_{(1)}\O$ is a pre-compact set, and therefore the sequence of
vectors $P_E\tU_n\O \in P_E\cA_{(1)}\O$ possesses a subsequence
$P_E\tU_{n(k)}\O$ converging strongly to $P_E\O^p$ (since
$\tU_n\O$ is already known to converge weakly to $\O^p$).
This allows us to write (with $\o^p(\,\cdot\,) = \langle \O^p,\,\cdot\,\O^p\rangle$)
\begin{eqnarray*}
\lefteqn{\frac{1}{i}\o^p(U_E^*\d(U_E)) = \o^p(F_E) = \o^p(P_EF_EP_E)}\\
& = & \lim_{k\to\infty}\,\langle P_E\tU_{n(k)}\O,F_EP_E\tU_{n(k)}\O\rangle\\
& = & \lim_{k\to\infty}\,\langle \tU_{n(k)}\O,F_E\tU_{n(k)}\O\rangle 
= \lim_{k\to
  \infty} \frac{1}{i}\o(\tU_{n(k)}^*U_E^*\d(U_E)\tU_{n(k)})\\
& = & \frac{1}{i}\lim_{k\to\infty}\,\{\o((U_E\tU_{n(k)})^*\d(U_E\tU_{n(k)})) -
\o(\tU^*_{n(k)}\d(\tU_{n(k)}))\} \\
&\ge& \liminf_{k\to\infty}\,\frac{1}{i}\o((U_E\tU_{n(k)})^*\d(U_E\tU_{n(k)})) -
\lim_{k\to\infty}\,\frac{1}{i}\o(\tU_{n(k)}^*\d(\tU_{n(k)}))\\
&\ge& c_{\o} - c_{\o} = 0\,.
\end{eqnarray*}
This shows that $\frac{1}{i}\o^p(U^*_E\d(U_E)) \ge 0$ for all $E > 0$,
and it is easy to check that also \\
 $\lim_{E \to \infty}\o^p(U^*_E\d(U_E)) =
\o^p(U^*\d(U))$. Therefore we conclude that
$$ \inf_{U\in\cU_0(\delta)}\frac{1}{i}\o^p(U^*\d(U)) \ge 0\,,$$
proving passivity of $\o^p$. \hfill $\Box$
\\[10pt]
Under the assumptions of Thm.\ \ref{P6} one finds,
by combining it with Thm.\ \ref{P3}, the following
%
\begin{Corollary}
Let $\O$ be a unit vector in $\cH$ which is energy compact, and
suppose that the corresponding vector state $\o$ fulfills a static
quantum weak energy inequality. Then there exists a unit vector in
$\overline{\cA\O}$ inducing a passive state for $\{\a_t\}_{t\in\bR}$. 
\end{Corollary}

\section{Some techniques from microlocal analysis}
\label{sect:uloc}

In the next section, we will show that the real scalar field on a static
spacetime is a system satisfying the assumptions made in the previous
section and to which our results therefore apply.  Our principal
tools will be drawn from microlocal analysis~\cite{Hormander1,Hfio1},
which provides powerful and geometrically natural techniques for dealing with the 
singular structure of distributions. We now proceed to describe the
microlocal analysis used in Sec.~\ref{sect:qfsb} and the Appendix,
adopting a slightly more intrinsic approach than usual, which avoids the
explicit introduction of coordinates. 

Let $X$ be a smooth manifold and denote by
$\dot{T}^*X=T^*X\backslash\cZ$ the cotangent bundle of $X$ with its zero
section $\cZ$ removed. Given a distribution $u\in\sD'(X)$, an element 
$(x,k)\in \dot{T}^*X$ is called a {\em regular directed
point} for $u$ if there exists a set $\cO$ and a map $\phi$ 
obeying\footnote{Condition (A) permits the introduction of coordinates
on $\cO$ by $y^\mu=\ip{\zeta^\mu}{\phi(y)}$ where $\zeta^\mu$ is any fixed
basis for $T_x^*X$, but such coordinates will rarely be necessary in our
discussion.}
\begin{itemize}
\item[(A)] $\cO$ is an open neighbourhood of $x$, and $\phi:\cO\to T_xX$ 
is a smooth map with nondegenerate tangent mapping $T\phi$ obeying $T\phi(x)={\rm
id}_{T_xX}$
\end{itemize}
and such that
\begin{itemize}
\item[(B)] there exists $\chi\in\Coin(\cO)$ with $\chi(x)\not=0$ and
a neighbourhood $E$ of $k$ in $T_x^*X$ such that 
\begin{equation}
\lambda^N \sup_{\ell\in E} \left| u(\chi e^{i\lambda\langle\ell,\phi\rangle})
\right|\to 0 \qquad \hbox{as $\lambda\to +\infty$ for each $N\in\NN_0$.}
\label{eq:WFintdef}
\end{equation}
\end{itemize}
The quantity inside the modulus signs can be
interpreted as a local Fourier transform\footnote{Throughout this paper,
we adopt the nonstandard convention $\widehat{f}(k)=\int
d^nx\,f(x)e^{ik\cdot x}$ for the Fourier transform on $\RR^n$.\label{fn:Fconv}}
of $\chi u$ evaluated at $\lambda\ell$: To see this, it may help to
write this quantity as
$$
u(\chi e^{i\lambda\langle\ell,\phi\rangle}) = \int d{\rm vol}(x) \chi(x)
u(x) e^{i\lambda\ell_a\phi^a(x)}\,,
$$ 
where $u(x)$ is the distributional kernel of $u$ with respect to
some volume measure $d{\rm vol}$ on $X$. 

As is well known, the Fourier transform of any smooth compactly
supported function decays rapidly at infinity, so every $(x,k)\in\dot{T}^*X$ is
regular directed for any distribution which may be identified with a
smooth function. It may also be shown that if $(x,k)$ is a regular directed point for $u$ then
condition~(B) will be satisfied for any pair $(\cO,\phi)$ obeying
condition~(A). Moreover, if condition~(B) holds, it continues to hold if $\chi$
is replaced by $\psi\chi$ for any $\psi\in\Coin(X)$. We are now in a
position to define the central object of the theory.
\begin{Definition} 
The {\em wave-front set} $\WF(u)$ of a distribution $u\in\sD(X)$ 
is the complement in $\dot{T}^*X$
of the set of regular directed points for $u$. 
\end{Definition}
Two important facts will be used extensively in this work: first, that
the wave-front set of a distribution $u$ is empty if and only if $u$ 
can be identified with a smooth function; second, if $P$ is a partial
differential operator with smooth coefficients then $\WF(Pu)\subset
\WF(u)$. 

Microlocal techniques have recently found many applications in the
theory of quantum fields on curved spacetimes following Radzikowski's
discovery~\cite{Rad1} that the Hadamard condition (see Sec.~\ref{sect:states})
can be reformulated as a condition on the wave-front set of the
two-point function of the field. Recently, this criterion has been 
simplified by Strohmaier, Wollenberg and Verch~\cite{SVW}, who consider a
generalisation of the wave-front set to distributions taking values in a
Hilbert space. Generalising this slightly further, if $(B,\|\cdot\|)$ is
a Banach space, let
$\sD'(X,B)$ be the space of distributions on $X$ taking values in $B$, 
i.e., linear functionals $T:\sD(X)\to B$
such that $f\to 0$ in $\sD(X)$ implies $\|T(f)\|\to 0$. Then the
wave-front set of $T$ may be defined as for scalar valued distributions,
but with the Banach space norm replacing the modulus signs
in~(\ref{eq:WFintdef}). With this definition, a convenient calculus may
be constructed as follows. 
\begin{Proposition} \label{prop:uloc}
Let $X$ be a $C^\infty$-manifold, $(\HH,\ip{\cdot}{\cdot})$ a Hilbert space and
$(B_i,\|\cdot\|_i)$ be Banach
spaces ($i=1,2$). \\
(i) If $T\in\sD'(X,B_1)$ and $S:\sD(X)\to B_2$ is a linear map obeying 
$$
\|S(f)\|_2\le c\|T(f)\|_1\,,\qquad f\in\sD(X)
$$
for some $c\ge 0$, then $S\in\sD'(X,B_2)$ and $\WF(S)\subset \WF(T)$. \\
(ii) If $T\in\sD'(X,\HH)$ and $\psi\in\HH$ then $f\mapsto\ip{\psi}{T(f)}$
and $f\mapsto\ip{T(\overline{f})}{\psi}$ define scalar distributions in $\sD'(X)$
with
$$
\WF(\ip{\psi}{T(\,\cdot\,)})\subset \WF(T(\,\cdot\,))\quad {\rm and}\quad
\WF(\ip{T(\overline{\,\cdot\,})}{\psi})\subset \WF(T(\,\cdot\,))^\dagger\,,
$$
where, for any $\Gamma\subset T^*X$, $\Gamma^\dagger=\{(x,k)\in T^*X\mid
(x,-k)\in\Gamma\}$. \\
(iii) If $S,T\in\sD'(X,\HH)$ then $U:(f,g)\mapsto
\ip{S(\overline{f})}{T(g)}$ defines $U\in\sD'(X\times X)$
with
$$
\WF(U)\subset \left(\WF(S)^\dagger\cup\cZ\right)\times 
\left(\cZ\cup\WF(T)\right)\,.
$$
\end{Proposition}
{\noindent\em Proof:} (i) The bound
$\|S(f)\|_2\le c\|T(f)\|_1$ implies that $S\in\sD'(X,B_2)$ and moreover that
any regular directed point for $T$ is a regular
directed point for $S$. The result follows on taking complements.\\
(ii) The statements regarding $f\mapsto\ip{\psi}{T(f)}$ follow
immediately from~(i) and the Cauchy-Schwarz inequality. To study
$f\mapsto\ip{T(\overline{f})}{\psi}$, it is convenient to prove an
auxiliary result first.
\begin{Lemma} \label{lem:conj}
Suppose $B$ is a Banach space equipped with a conjugation
$\Gamma$ and that $S\in\sD'(X,B)$. Defining
\[
S^\dagger(f) = \Gamma S(\overline{f})\,,\qquad f\in\sD(X)\,,
\]
we have $S^\dagger\in\sD'(X,B)$ and $\WF(S^\dagger)=\WF(S)^\dagger$.
\end{Lemma}
{\noindent\em Proof:} Since $\|S^\dagger(f)\| = \|S(\overline{f})\|$ and
$f\to 0$ if and only if $\overline{f}\to 0$, we have
$S^\dagger\in\sD'(X,B)$. Furthermore, it is easy to see that $(x,k)$ is a regular direction
for $S$, $(x,-k)$ is a regular direction for $S^\dagger$, so
$\WF(S^\dagger)\subset\WF(S)^\dagger$; since $(S^\dagger)^\dagger=S$, we
must in fact have equality.\hfill $\Box$

Now any Hilbert space admits a conjugation, and the Cauchy-Schwarz
inequality gives
\[
|\ip{T(\overline{f})}{\psi}| \le
\|\psi\|\,\|T(\overline{f})\|=\|\psi\|\,\|T^\dagger(f)\|\,,
\]
so we apply~(i) and Lemma~\ref{lem:conj} to complete the proof of~(ii).\\
(iii) Applying Cauchy-Schwarz, 
\[
|U(f,g)| \le \|S(\overline{f})\|\,\|T(g)\| =\|S(f)^\dagger\|\,\|T(g)\| =
\| (S^\dagger\otimes T)(f,g)\| \,,
\]
where $S^\dagger\otimes T\in\sD'(X\times X,\HH\otimes\HH)$ is defined by
$(S^\dagger\otimes T)(f,g)=S^\dagger(f)\otimes T(g)$. The same arguments
which bound the wave-front set of a tensor product of scalar
distributions may be used to show that 
\[
\WF(S^\dagger\otimes T)\subset \left(\WF(S^\dagger)\cup\cZ\right)\times
\left(\cZ\cup\WF(T)\right)
\]
[actually, there is a tighter bound than this]; using~(i) and Lemma~\ref{lem:conj}
the required result is obtained. \hfill $\Box$

A further important property required below is the behaviour of distributions under the
pull-back operation.
Let $X_1$ and $X_2$ be smooth manifolds and let $\chi: X_1 \to X_2$ be
a $C^\infty$-map. To this map one can associate its {\em conormal bundle} 
$N_\chi \subset T^*X_2$ where, by definition, $(y,\eta)$ is
in $N_\chi$ if and only if there is $x \in X_1$ with $y = \chi(x)$ and
${}^tT\chi(x)\eta = 0$, ${}^tT\chi(x)$ denoting the transpose of the
tangent map of $\chi$ at $x$.

If $F : X_2 \to \mathbb{C}$ is a smooth map, one obtains via the
pull-back by $\chi$ a smooth map $\chi^*F = F \circ \chi: X_1 \to
\mathbb{C}$. It can be shown (cf.\ Thm.\ 8.2.4 in \cite{Hormander1})
that one can (uniquely) extend the pull-back operation to
distributions $u \in \sD'(X_2)$ --- through approximating
distributions by test functions --- provided that
$$ N_{\chi} \cap \WF(u) = \emptyset\,.$$
In this case, the pull-back $\chi^*u$ of $u$ by $\chi$ has the
property
$$ \WF(\chi^*u) = \chi^*\WF(u)\,,$$
where for any subset $V \subset T^*X_2$, $\chi^*V$ is the subset of
$T^*X_1$ defined as follows:
$$ \chi^*V = \{(x,{}^tT\chi(x)\eta): (f(x),\eta) \in V\}\,.$$
Moreover (see again Thm.\ 8.2.4 in \cite{Hormander1}), $\chi^*$
induces a continuous linear map from $\sD'_V(X_2)$ into
$\sD'_{\chi^*V}(X_1)$ if $V \cap N_{\chi}= \emptyset$. Here, for each
closed conic subset $V \subset T^*X_2$, the set $\sD'_{V}(X_2)$ is a
linear subspace of the distribution space $\sD'(X_2)$ which is defined
as $\sD'_V(X_2) = \{u \in \sD'(X_2): \WF(u) \subset V\}$. As we will
discuss below, there is a
notion of convergence in $\sD'_V(X_2)$ with respect to which
$\sD'_V(X_2)$ is a closed subset of $\sD'(X_2)$ (cf.\ also Def.\ 8.2.2 in
\cite{Hormander1}). This notion of convergence is sometimes referred to
as the ``H\"ormander pseudo-topology'' of $\sD'_V(X_2)$ and it is this
sense in which $\chi^*$ is continuous, $\sD'_{\chi^*V}(X_1)$ being
analogously defined.

Convergence in the H\"ormander pseudo-topology may be defined as
follows. Suppose $u_r$ is a sequence in $\sD'_V(X)$ and $u\in\sD'_V(X)$.
Then $u_r\to u$ in $\sD'_V(X)$ if the two following conditions hold:
\begin{itemize}
\item[(i)] $u_r\to u$ weakly in $\sD'(X)$ (i.e., $u_r(f)\to u(f)$ for 
each test function $f$)
\item[(ii)] for all $(x,k)\in \dot{T}^*X\backslash V$, there exists
$(\cO,\phi)$ obeying condition~(A) above, $\chi\in\Coin(\cO)$ with
$\chi(x)\not=0$ and a neighbourhood $E$ of $k$ in $T_x^* X$ obeying
$$
\left[\phi^*\left(\phi(\supp \chi)\times E\right)\right] \cap V=\emptyset
$$
and such that the quantities
$$
\sup_{\lambda\in\RR^+} \sup_{\ell\in E} \lambda^N |u_r(\chi
e^{i\lambda\ip{\ell}{\phi}})| 
$$ 
are uniformly bounded in $r$ for each $N=1,2,\ldots$. 
\end{itemize}
Any distribution $u\in\sD'(X)$ may be arbitrarily well approximated in $\sD'_V(X)$, for
any $V\supset\WF(u)$, by a sequence of test functions $u_r\in\sD(X)$; it
is this which permits the definition of pull-backs to be extended to
distributions by continuity from the definition for functions. 

The notion of H\"ormander pseudo-topology is easily extended to distributions taking
values in a Banach space $B$ simply by replacing modulus signs with
Banach norms where appropriate, and denoting by $\sD'_V(X,B)$ the set of
distributions in $\sD'(X,B)$ whose wave-front sets are contained in $V$.

It is useful to have some simpler sufficient conditions for
convergence in $\DD_V'(X,B)$. It is not hard to show that, for
example, if $u_r$ is a sequence converging weakly to $u$ in $\sD'(X,B)$
and $\|u_r(f)\|\le \|v(f)\|$ for all $r$ and some
$v\in\sD'(X,B)$ then $u_r\to u$ in $\sD_{\WF(v)}'(X)$. The following is
a slight elaboration of this observation.
\begin{Proposition} \label{prop:Hpsdom}
Let $X$ and $Y$ be $C^\infty$-manifolds and $(B_i,\|\cdot\|_i)$ be
Banach spaces ($i=1,2$). Suppose that $u_r$ is a sequence in
$\sD'(X\times Y,B_1)$ converging weakly to $u$. Suppose further that
there exists $v\in\sD'(X\times Y,B_2)$ such that
\begin{equation}
\label{eq:urvbd}
\|u_r(f,g)\|_1\le \|v(f,g)\|_2 
\end{equation}
for all $f\in\sD(X)$ and $g\in\sD(Y)$. Then $u_r\to u$ in the
H\"ormander pseudo-topology on $\sD'_{\WF(v)}(X\times Y)$.
\end{Proposition}
{\noindent\em Remark:} The result continues to hold if
Eq.~(\ref{eq:urvbd}) is generalised to
\[
\|u_r(f,g)\|_1\le \sum_{i=1}^n\|v_i(f,g)\|_i 
\]
for Banach spaces $(B_i,\|\cdot\|_i)$ and distributions
$v_i\in\DD'(X\times Y,B_i)$, ($i=1,\ldots, n$) by applying the
proposition to $v=v_1\oplus v_2\oplus\cdots\oplus v_n\in\DD'(X\times Y, 
\bigoplus_{i=1}^n B_i)$. \\
{\noindent\em Proof:} Suppose $(x_0,y_0;k_0,l_0)$ is a regular directed
point for $v$. Let $(\cO_X,\phi_X)$ and $(\cO_Y,\phi_Y)$ obey
condition~(A) above for points $x_0\in X$ and $y_0\in Y$, and define
$$
\phi(x,y)=\phi_X(x)\oplus\phi_Y(y)\in T_{x_0}X\oplus T_{y_0}Y \cong 
T_{(x_0,y_0)} (X\times Y)\,.
$$
Then $(\cO_X\times \cO_Y, \phi)$ obeys condition~(A) for $(x_0,y_0)$;
furthermore, by continuity of the pull-back $\phi^*$ and the fact that
$\WF(v)$ is closed, there exists open $\cO\subset \cO_X\times \cO_Y$ and
a neighbourhood $G$ of $(k_0,l_0)$ in $T^*_{(x_0,y_0)}X\times Y$ such
that
$$
\left[\phi^*\left(\phi(\cO)\times G\right)\right]\cap \WF(v)=\emptyset\,.
$$
Since $(x_0,y_0;k_0,l_0)$ is regular directed for $v$, there exists
$\chi\in\Coin(\cO)$ with $\chi(x_0,y_0)\not=0$ and a neighbourhood $E$
of $(k_0,l_0)$, contained (without loss of generality) in $G$ such that
\begin{equation}
\lambda^N \sup_{(k,l)\in E}\left\|v\left(\chi
e^{i\lambda\ip{(k,l)}{\phi}}\right)\right\|_2 \to 0\qquad {\rm as}~\lambda\to
+\infty
\label{eq:vbnd}
\end{equation}
for each $N$. Choose smooth functions $\eta_X$ and $\eta_Y$ such that
$\eta(x,y)=\eta_X(x)\eta_Y(y)$ is compactly supported in the interior of
$\supp \chi$, with $\eta(x_0,y_0)\not=0$. Then 
$$
\psi(x,y) = \left\{\begin{array}{cl} \eta(x,y)/\chi(x,y) &
(x,y)\in\supp\eta\\
0 & \hbox{otherwise} \end{array}\right.
$$
is smooth and compactly supported and Eq.~(\ref{eq:vbnd}) continues to
hold if $\chi$ is replaced by $\psi\chi=\eta$, thereby yielding
$$
\lambda^N \sup_{(k,l)\in E}\left\|v\left(\eta_X
e^{i\lambda\ip{k}{\phi_X}},\eta_Y
e^{i\lambda\ip{l}{\phi_Y}}\right)\right\|_2 \to 0\qquad {\rm as}~\lambda\to
+\infty
$$
for each $N$. Together with Eq.~(\ref{eq:urvbd}) this immediately
implies that $(x_0,y_0;k_0,l_0)$ is a regular directed
point for each $u_r$ and $u$. It follows that $u_r$ and $u$ belong to
$\sD'_{\WF(v)}(X,B_1)$. Furthermore, 
$$
\sup_{\lambda\in\RR^+}\sup_{(k,l)\in E}
\lambda^N \left\|u_r(\chi
e^{i\lambda\ip{\ell}{\phi}})\right\|_1
\le  \sup_{\lambda\in\RR^+}\sup_{(k,l)\in E}
\lambda^N
\left\|v\left(\eta_X
e^{i\lambda\ip{k}{\phi_X}},\eta_Y
e^{i\lambda\ip{l}{\phi_Y}}\right)\right\|_2\,,
$$
the right-hand side of which is easily seen to be finite. This provides the required
uniform bound to ensure that $u_r\to u$ in $\sD'_{\WF(v)}(X,B_1)$. 
\hfill $\Box$

\section{Quantum fields on static backgrounds}
\label{sect:qfsb}
We will now describe how the structural assumptions made in
Sec.~\ref{sect:absres} may be justified for the case of real scalar
field theory on a 
globally hyperbolic static spacetime $(M,\gb)$ with compact spatial
sections. The assumptions to be checked are:
\begin{itemize}
\item the existence of a $C^*$-dynamical system along with a suitable 
sub-$*$-algebra $\cA^\infty$ and a 
generating linear space $\cW^\infty$ which is stable under convolutions;
\item the identification of a set of states $\cS$ closed under finite convex
combinations and operations induced by elements of the algebra
 $\cU^\infty_{\rm alg}$ constructed from $\cA^\infty$;  
\item the existence of an energy density [defined for every state in $\cS$]
whose spatial integral generates the dynamics;
\item the existence of states satisfying a suitable static QWEI.
\end{itemize}
Each assumption will be treated in turn in the following subsections. 
Most details are postponed to the Appendix.
\\[6pt]
It is worth mentioning that we will also prove a converse to 
Thm.~\ref{P5}(a) for the free scalar field: namely, we will show in
Thm.~\ref{Thm:GSQ} 
that a non-degenerate ground state with mass gap and vanishing one-point
functions is necessarily quiescent. 
\subsection{The dynamical system} \label{sect:dynsys}

We begin by reviewing the quantisation of the real scalar field on a
globally hyperbolic static spacetime $(M,\gb)$.
Such a spacetime is diffeomorphic to $\RR\times \Sigma$ with 
line element
$$
ds^2 = g_{ab} dx^a dx^b = g_{00}dt^2 - h_{ij} dx^i dx^j\,,
$$
where $\hb$ is a (positive definite) Riemannian metric and  
$g_{00}$ is a smooth strictly positive function on $\Sigma$. 
As before, we will assume that $\Sigma$ is $s$-dimensional and compact;
the preferred measure on $\Sigma$ is $d\mu(\ux) = \sqrt{h}d^s\ux$, where
$h=\det\hb$. The Killing vector $\partial/\partial t$ will be denoted $\xi$.
We will also introduce an orthonormal frame $e_\mu^a$ ($\mu=0,\ldots,s$) with
$e_0^a=g_{00}^{-1/2}\xi^a$.

The Klein--Gordon equation on $(M,\gb)$ is
$$
\left(g^{ab}\nabla_a\nabla_b + m^2\right)\varphi =0\,,
$$
for which the corresponding classical stress-energy tensor is
$$
T_{ab} = \nabla_a \varphi \nabla_b \varphi - \frac{1}{2}g_{ab}g^{cd}
\nabla_c \varphi \nabla_d \varphi + \frac{1}{2}m^2 g_{ab} \varphi^2\,.
$$
Integrating over a surface of constant `time' $\{t\}\times\Sigma$ 
($t\in\RR$) we obtain the classical energy
$$
H = \int_\Sigma T_{ab}(t,\ux) n^a\xi^b d\mu(\ux)\,,
$$
where $n^a=g_{00}^{-1/2}\xi^a=e_0^a$ is the future-pointing unit normal to
$\Sigma$; this is conserved by virtue of the Klein--Gordon equation and
Gauss' theorem.  In addition, the classical energy density seen by an
observer with velocity $e_0^a$ is 
\begin{equation}
e_0^a e_0^b T_{ab} = \frac{1}{2}\left(m^2\varphi^2 + \sum_{\mu=0}^s
(e_\mu^a\nabla_a\varphi)^2\right) \,.
\label{eq:cled}
\end{equation}

The quantisation of this system proceeds as follows (cf.\ \cite{Dim.KG}): 
first, let $(S,\sigma)$ be the symplectic space of
smooth real-valued Klein--Gordon solutions
 where the symplectic form is given by
$$
\sigma(u,v) = \int_\Sigma d\mu(\ux) \left(u e_0^a\nabla_a v -
ve_0^a\nabla_a u \right)\,.
$$
The CCR-algebra (or Weyl algebra) $\fA[S,\sigma]$ over $(S,\sigma)$
is the (unique up to $C^*$-isomorphism) 
unital $C^*$-algebra generated over $\CC$ by unitary elements 
${\sf W}(u)$ ($u\in S$) with ${\sf W}(0)={\bf 1}$ subject to the Weyl relations
$$
{\sf W}(u) {\sf W}(v) = e^{-i\sigma(u,v)/2} {\sf W}(u+v)\,, \qquad u,v\in S\,.
$$

For each open relatively compact $O\subset M$, let $\fA(O)$ 
be the sub-$C^*$-algebra of $\fA[S,\sigma]$ generated by elements of the form
${\sf W}(Ef)$ where $f\in \Coin(O;\RR)$ and $E:\Coin(M)\to \Cin(M)$ is the
advanced-minus-retarded fundamental solution for the Klein--Gordon
equation. Then $O\mapsto \fA(O)$ is an
isotonous net of $C^*$-algebras which is also local in
the sense that $A_1A_2 = A_2A_1$ holds for all $A_1 \in \fA(O_1)$ and
$A_2 \in \fA(O_2)$ whenever the regions $O_1$ and $O_2$ in $M$ cannot
be connected by any causal curve. Hence, the net $O\mapsto \fA(O)$ is
the essential building block of a local quantum field theory on the
curved spacetime $(M,\gb)$ (cf.\ \cite{Haag,WaldII}).

Since the time translations $\tau_{t}(t',\ux)=(t+t',\ux)$ induce a
symplectomorphism of $(S,\sigma)$ there is a 1-parameter group of
$*$-automorphisms on $\fA[S,\sigma]$ given by
$$
\aaa_t ({\sf W}(u)) = {\sf W}(\tst u)\,, 
\qquad u\in S\,, \ t \in \bR\,,
$$
where $\tst u=u\circ\tau_t^{-1}$ is the push-forward.\footnote{The
push-forward $\tst$ and pull-back $\tau^*_t$ are related by
$\tau_{-t}^*u = u \circ (\tau_{-t}) = u\circ \tau_t^{-1} =
\tau_{t}{}_*u$.} The
$C^*$-algebraic net $O\mapsto \fA(O)$ then has the covariance
 property~\cite{Dim.KG}
$$ 
\aaa_t(\fA(O)) = \fA(\t_t(O))\,. 
$$
However, $\{\aaa_t\}_{t\in\RR}$ is not strongly continuous (because
$\|{\sf W}(u)-{\sf W}(v)\|=2$ for all $u\not=v$). This obstacle can be
circumvented
as we shall explain, but let us first 
give a brief description of how we proceed. 
We will, in the following subsection, work in GNS-representations of quasifree
Hadamard states which may be regarded as states in quantum field
theory on curved spacetimes 
whose short-distance behaviour is close to that of vacuum states or thermal
equilibrium states. We shall denote by ${\cal S}_{\sf{q H}}$
the set of all quasifree Hadamard states on $\fA[S,\sigma]$, and
starting from this class we shall define the underlying
$C^*$-dynamical system, the sub-$*$-algebra $\cA^{\infty}$ and the set
of states $\cal S$. In the remainder of this subsection we will discuss
the $C^*$-dynamical system, and the algebra $\cA^{\infty}$.

Consider a state $\omega$ on $\fA[S,\sigma]$ and let
$(\cH_{\o},\pi_{\o},\O_{\o})$ be its GNS-representation. Then we call
such a state {\it weakly covariant} if there exists on the von Neumann
algebra $\cM_{\o} = \pi_{\o}(\fA[S,\sigma])''$ a one-parameter group
$\{\alpha^{(\o)}_t\}_{t\in\bR}$ of automorphisms (leaving $\cM_{\o}$
invariant) so that 
\begin{equation} \label{covprop}
\a^{(\o)}_t \circ \pi_{\o}({\sf A}) = \pi_{\omega} \circ 
\aaa_t({\sf A})\,, \quad
t \in \bR,\ {\sf A} \in \fA[S,\sigma]\,.
\end{equation}
A special case is a covariant state $\o$ where $\alpha^{(\o)}_t(A) =
V_t^{(\o)}AV_t^{(\o)}{}^*$ with a strongly continuous unitary group
$\{V_t^{(\o)}\}_{t\in\bR}$ on $\cH_{\o}$. Then
$(\cM_{\o},\{\a^{(\o)}_t\}_{t\in\bR})$ is a $W^*$-dynamical system (cf.\
\cite{BR1}), but we need to pass to a $C^*$-dynamical
system whose definition we now describe, following the strategy
outlined in the Introduction. Consider the operators $\a_fA$
defined by \eqref{convdef} with $\a_t \equiv \a_t^{(\o)}$ for $f \in
C^{\infty}_0(\bR)$ and $A\in \pi_{\o}(\fA[S,\sigma])$ where the
integral is understood in the weak topology, so $\a_fA \in
\cM_{\o}$. It is straightforward to check that $||\a_fA|| \le
||f||_{L^1}||A||$ and $\a^{(\o)}_t(\a_fA) = \a_{f(\,.\,-t)}A$, so that
$||\a^{(\o)}_t(\a_fA) -\a_fA|| \to 0$ for $t\to 0$. Now we define
$\cA_{\o} \subset \cM_{\o}$ as the $C^*$-closure of the $*$-algebra
generated by all these $\a_fA$; then
$(\cA_{\o},\{\a_t^{(\o)}\}_{t\in\bR})$ is a $C^*$-dynamical system
(where $\a^{(\o)}_t$ should here strictly be read as $\a^{(\o)}_t\rest
\cA_{\o}$). Now this dynamical system depends on the chosen state
$\o$, but for our discussion, this dependence is
spurious, as we shall explain. 

As mentioned above, we denote by ${\cal S}_{\sf{q H}}$ the set of
quasifree Hadamard states on $\fA[S,\sigma]$ (see next subsection for
a definition). At the present stage of discussion it is important to
know that, whenever $\o_1$ and $\o_2$ are contained in ${\cal
  S}_{\sf{q H}}$, then $\o_1$ and $\o_2$ are quasi-equivalent
\cite{Ver1} (because of spatial compactness of the underlying
spacetime).
 This means in the notation just introduced that there are
von Neumann-algebra isomorphisms $\b_{21}:\cM_{\o_1} \to \cM_{\o_2}$
and $\b_{12}: \cM_{\o_2} \to \cM_{\o_1}$ such that $\b_{21} \circ
\pi_{\o_1} = \pi_{\o_2}$ and $\b_{12} \circ \pi_{\o_2} = \pi_{\o_1}$. We
recall also that all quasifree states on $\fA[S,\sigma]$ are faithful,
and so are their GNS-representations. Thus
$\b_{21}=\b_{12}^{-1}$. Moreover, if any state $\o_1 \in {\cal
  S}_{\sf{q H}}$ turns out to be covariant, then all $\o_2 \in
{\cal S}_{\sf{q H}}$ are weakly covariant, too, with
$$ 
\a^{(\o_2)}_t = \b_{21} \circ \a_t^{(\o_1)} \circ \b_{12}\,.
$$
It is also straightforward to check that $\cA_{\o_2} =
\b_{21}(\cA_{\o_1})$. In this sense, the $C^*$-dynamical system
$(\cA_{\o}, \{\a^{(\o)}_t\}_{t\in \bR})$ is independent of the chosen
state $\o \in {\cal S}_{\sf{q H}}$ once it is known that there
exist quasifree Hadamard states for $\fA[S,\sigma]$ which are weakly
covariant. This, however,
can also be concluded from the fact that each pair of states in ${\cal
  S}_{\sf qH}$ is quasi-equivalent. To see this one simply notes that
for each $\o \in {\cal S}_{\sf qH}$ the time-shifted state $\o \circ
\a_t$ is again in ${\cal S}_{\sf qH}$ -- this is a consequence of the
fact that the wave-front set of the two-point function (cf.\ Sec.\
4.2) of a quasifree Hadamard state is left invariant under the
isometries $\t_t$. Thus, since $\pi_{\o} \circ \tilde{\a}_t$ and $\pi_{\o
  \circ \tilde{\a}_t}$ are canonically unitarily equivalent by the uniqueness
of the GNS-representation, there is for each $t$ a von Neumann
algebraic isomorphism $\a^{(\o)}_t: \cM_{\o} \to \cM_{\o}$ with the
covariance property \eqref{covprop}.
 Hence, starting from the class of states ${\cal 
  S}_{\sf{q H}}$, we see that there is (up to isomorphism) a
unique $C^*$-dynamical system associated with it. 

Now, for $\o \in {\cal S}_{\sf{q H}}$ we define the sub-vector space
$\cW^{\infty}_{\o}$ of $\cA_{\o}$ as the vector space generated by all 
$\a_f\pi_{\o}({\sf W}(u))$ where $f \in C_0^{\infty}(\bR)$
and $u \in S$, and denote by $\cA^\infty_{\omega}$ the $*$-algebra
generated by $\cW_{\omega}^\infty$. As a consequence of the
Weyl-relations, it is straightforward to check that
$\cW_{\omega}^\infty$ is norm-dense in $\cA_{\omega}$ and stable under
taking adjoints; it is stable under convolution with test-functions
with respect to $\alpha_t$ by its very definition. Moreover, one
may easily check that $\cA^{\infty}_{\o_2} =
\b_{21}(\cA^{\infty}_{\o_1})$ for all $\o_1,\o_2 \in {\cal
  S}_{\sf{q H}}$. The vector space $\cW^\infty_{\omega}$ and algebra
$\cA^{\infty}_{\o}$ 
are uniquely associated with ${\cal S}_{\sf{q H}}$ up to isomorphism,
just as the $C^*$-dynamical system was. In this
light, we shall 
henceforth adopt the following conventions:
\begin{itemize}
\item[$-$] \quad we choose an arbitrary, quasifree Hadamard state $\om_0$
  and keep it fixed,
\item[$-$] \quad we denote by $(\cA,\{\a_t\}_{t\in\bR})$, $\cW^\infty$ 
  and $\cA^{\infty}$ the $C^*$-dynamical system, dense  \\
 ${}$ \quad subspace and $*$-algebra 
  associated with $\om_0$ as just described.
\end{itemize}
In applying the abstract results of Sec.~\ref{sect:absres}, we will take
$(\cA,\{\alpha_t\}_{t\in\RR})$ to be the $C^*$-dynamical
system of interest. 

The results of Sec.~\ref{sect:absres} then assert the existence of
passive states $\o^p$ on $\cA$ as a consequence of suitable forms of
static quantum weak energy inequalities (which will be established in the
following sections) and the question might arise under which
conditions the states $\o^p$ can be interpreted as passive states on
the original Weyl algebra $\fA[S,\sigma]$. The following lemma shows
that these states always induce states on $\fA[S,\sigma]$ (recalling
that a passive state is always  invariant under the time-evolution).
\begin{Lemma} Let $\o$ be an $\{\a_t\}_{t\in\bR}$-invariant state on
  $\cA$. Then $\o$ induces a $\{\tilde{\a}_t\}_{t\in\bR}$-invariant
  state ${\o}^{\fA}$ on $\fA[S,\sigma]$.
\end{Lemma}
{\it Proof. } For all $f \in C_0^\infty( \bR)$ and all $A =
\pi_{\o_0}({\sf A}) \in \pi_{\o_0}(\fA[S,\sigma])$ the
estimate
$$ |\o(\a_fA)| \le ||f||_{L^1}||A|| $$
entails that, for fixed $A$, $f \mapsto \o(\a_fA)$ extends to a continuous
linear functional on $L^1(\bR)$. Consequently, there exists a function
$L_A \in L^\infty(\bR)$ so that
$$ \o(\a_fA) = \int dt\,L_A(t)f(t)\,, \quad f \in L^1(\bR)\,.$$
Now as $\o$ is $\{\a_t\}_{t\in\bR}$-invariant, it follows easily that
$L_A$ must be constant (almost everywhere). Let us denote this
constant by $\overline{\o}(A)$, then it holds that
$$ \o(\a_fA) = \overline{\o}(A)\int dt\,f(t)\,, \quad f \in
L^1(\bR)\,,$$
showing that $\overline{\o}(A) = \o(\a_fA)$ whenever $\int dt \,f(t) =
1$\,. The assignment $A \mapsto \overline{\o}(A)$ is obviously linear
and we need to show that it fulfills state-positivity. Let $B = A^*A$
be a positive element in $\pi_{\o_0}(\fA[S,\sigma])$. All we need to
demonstrate is the existence of some $f \in L^1(\bR)$ with $\int dt\,
f(t) = 1$ so that $\a_fB$ is a positive element in $\cA$ (whereupon
$\overline{\o}(B) = \o(\a_fB) \ge 0$). Choosing any $f \ge 0$ in
$L^1(\bR)$ with $\int dt\,f(t) = 1$, it is clear that $\a_fB$ is a
positive element in $\cB(\cH_{\o_0})$. But since $\cA$ inherits the
$*$-operation and $C^*$-norm of $\cB(\cH_{\o_0})$, it follows that
$\a_fB$ is also a positive element in $\cA$ (cf.\ Lemma 2.2.9 in
\cite{BR1}). Moreover, $\overline{\o}(\a_t(A)) = \o(\a_f(\a_t A))=\o(\a_t(\a_fA)) =
\o(\a_fA) = \overline{\o}(A)$ so $\overline{\o}$ is an
$\{\a_t\}_{t\in\bR}$-invariant state on $\pi_{\o_0}(\fA[S,\sigma])$.
Thus $\o^{\fA} = \overline{\o} \circ \pi_{\o_0}$ is an
$\{\tilde{\a}_t\}_{t\in\bR}$-invariant state on $\fA[S,\sigma]$.
${}$ \quad ${}$ \hfill $\Box$
\\[10pt]
However, it should be noted that one has no information regarding the
continuity of $\o^{\fA}$ with respect to the time-evolution, in other
words, there is no reason why the functions $t \mapsto \o^{\fA}({\sf
  A}\tilde{\a}_t({\sf B}))$, ${\sf A,B} \in \fA[S,\sigma]$, should be
continuous, and therefore it is unclear if $\o^{\fA}$ is passive (in a
$W^*$-sense) on $\fA[S,\sigma]$  if
$\o$ is a passive state on $\cA$. This can be concluded if $\o$
fulfills further regularity conditions. A sufficient condition to that
effect is that
$\o = \o^p$ be a normal state on $\cM_{\o_0}$, and we have seen in
Cor.\ 2.7 that a certain energy-compactness condition ensures this normality.

We should also like to point out that one can generalize the notion of
$n$-point correlation functions so that it is applicable to states on
$\cA$ in the sense that sufficently regular states
(``$C^\infty$-regular states'') on $\cA$ possess
$n$-point correlation functions for all $n \in \bN$, inducing states on
the algebra of abstract Klein-Gordon field operators. These matters
will be discussed in Appendix \ref{sect:corell}. The result of this
discussion
again shows that there is hardly any difference in working with
$\fA[S,\sigma]$ or $\cA$ as long as ``sufficiently regular'' states
are considered, and thus our passing from $\fA[S,\sigma]$ to $\cA$ can
rightfully be regarded as made purely for technical convenience.

\subsection{The state space $\cS$}
\label{sect:states}

The states to be considered are drawn from the class of {\em Hadamard}
states on $\fA[S,\sigma]$, for which the renormalised energy density
may be defined by point-splitting. They are defined as follows.
Suppose a
state $\omega$ is sufficiently regular that the function $(s,t)\mapsto
\om({\sf W}(sEf){\sf W}(tEg))$ is twice continuously 
differentiable for each pair of
real-valued test functions $f,g$ and that, moreover, the two-point
function $w_2^{(\o)}$ defined by
$$
w_2^{(\o)}(f,g)=-\left.\frac{\partial^2}{\partial s\partial t}  
\omega({\sf W}(sEf){\sf W}(tEg))\right|_{s,t=0}\,,\qquad f,g\in\sD(M;\RR)
$$
extends (by complex linearity in its arguments) to a distribution in
$\sD'(M\times M)$. Then $\om$ is said to be Hadamard if the corresponding 2-point
correlation function $w_2^{(\o)}$ takes
the so-called Hadamard form~\cite{KayWald}, which 
completely fixes $w_2^{(\o)}$ modulo smooth
terms; in particular, the difference between any two Hadamard two-point
functions is smooth. In~\cite{Rad1}, Radzikowski showed that this
condition could be replaced by the requirement that the wave-front set
of the two-point function should satisfy
\begin{equation}
\WF(w_2^{(\o)}) = \{(x,k;x',-k')\in \dot{T}^*(M\times M): (x,k)\sim
(x',k'),~k\in\cN_x^+\}\,,
\label{eq:RadHad}
\end{equation}
where $\cN_x^+$ is the cone of (non-zero) future-pointing null covectors
at $x$ and $(x,k)\sim (x',k')$ if there is a null geodesic connecting
$x$ and $x'$ to which $k$ and $k'$ are cotangent at $x$ and $x'$
respectively, with $k'$ being the parallel transport of $k$. 
[In the case $x=x'$, we require $k=k'$.] For future
reference we will also use $\cN_x^-$ to denote the cone of non-zero
past-pointing null covectors at
$x$ and $\cN^\pm=\bigcup_{x\in M} \cN_x^\pm$ for the future and past
null cones in $T^*M$.

Radzikowski's criterion has been simplified recently by 
Strohmaier, Wollenberg and Verch~\cite{SVW}, who consider
Hilbert-space valued distributions induced by the field operators. Their
characterisation is essentially the following.
\begin{Theorem} \label{Thm:SVW}
A state $\om$ on $\fA[S,\sigma]$ is Hadamard if and only
if the following conditions hold in some GNS representation [not
necessarily that induced by $\om$] $(\HH,\pi,\Omega)$ of $\fA[S,\sigma]$:
\renewcommand{\theenumi}{\alph{enumi})}
\begin{enumerate}
\item $\om$ is represented by a vector $\psi\in\HH$, i.e.,
$\om(A)=\ip{\psi}{\pi(A)\psi}$ for all $A\in\fA[S,\sigma]$;
\item the function $t\mapsto \pi({\sf W}(tEf))\psi$ is differentiable for all
$f\in\sD(M;\RR)$;
\item the $\HH$-valued functional $f\mapsto \Phi(f)\psi:=-id/dt\,
  \pi({\sf W}(tEf))\psi|_{t=0}$ 
extends by complex-linearity to a Hilbert-space valued distribution
$\Phi(\cdot)\psi\in\sD'(M,\HH)$ obeying
\begin{equation}
\WF(\Phi(\cdot)\psi)\subset\cN^-\,.
\label{eq:Hadamard}
\end{equation}
\end{enumerate}
\end{Theorem}
{\noindent\it Remark:} Condition {\it a)} may always be satisfied by using the
GNS representation induced by $\om$, but it is convenient to allow for
other representations. Note that the assignment $f \mapsto \Phi(f)\psi$
defines the field operator $\Phi(f)$ in the GNS-Hilbert-space
representation $(\mathcal{H},\pi,\Omega)$ on the domain $D(\Phi(f))$
of all $\psi \in \mathcal{H}$ for which $-i d/dt \pi({\sf
  W}(tEf))\psi|_{t=0}$ exists as strong limit in $\mathcal{H}$.
\\[10pt]
Now let $\o$ be any quasifree Hadamard state state on $\fA[S,\sigma]$,
and denote by $(\cH_{\om},\pi_{\om},\Omega_{\om})$ the corresponding
GNS-representation, and by $\Phi_{\om}(f) = -id/dt\pi_{\om}({\sf
  W}(tEf))|_{t=0}$ the field operators, defined on a dense domain
$D(\Phi_{\om}(f)) \subset \cH_{\om}$ for $f \in \mathscr{D}(M)$. We
define ${\sf Had}(\om)$ as the set of vectors $\psi\in\bigcap_{f \in
  \mathscr{D}(M)} D(\Phi_{\om}(f))$ having the property that
$\Phi_{\om}(\cdot)\psi$ belongs to $\mathscr{D}'(M,\cH_{\om})$ and obeys the
Hadamard condition 
\eqref{eq:Hadamard}. For every quasifree Hadamard state $\om$, the
vectors $\psi \in \Had(\om)$ induce states and, more generally,
continuous linear functionals on the $C^*$-algebra $\cA=\cA_{\om_0}$
of our dynamical system. Namely, let $\beta_{\om\om_0}:
\mathcal{M}_{\om_0} \to \mathcal{M}_{\om}$ be the von Neumann-algebra
isomorphism with $\beta_{\om\om_0} \circ \pi_{\om_0} = \pi_{\om}$, then
$$ \om^{[\psi]}(A) = \langle \psi,\beta_{\om\om_0}(A)\psi \rangle\,,
\quad A \in \cA_{\om_0}\,,$$
is a state on $\cA = \cA_{\om_0}$. We now define the state space $\cS$
as the set of finite convex combinations of states induced by vectors
in $\Had(\om)$, $\om \in \cS_{\sf qH}$. In other words, a state
$\tilde{\om}$ on $\cA$ is contained in $\cS$ iff there are finitely
many quasifree Hadamard states $\om_i \in \cS_{\sf qH}$ ($i=
1,\ldots,N$) together with unit vectors $\psi_i \in \Had(\om_i)$ and
$\lambda_i > 0$, $\sum_{i=1}^N \lambda_i = 1$, such that
$$ \tilde{\om}(A) = \sum_{i=1}^N \lambda_i \om_i^{[\psi_i]}(A)\,,
\quad A \in \cA\,.$$ 
Theorem~\ref{Thm:SVW} guarantees that all the states
in $\cS$ are Hadamard states. The state space $\cS$ has the following
properties, as will be proved in Appendix A.1:
\begin{Proposition} \label{Prop:stability}
$\cS_{\sf qH} \subset \cS$; and 
$\cS$ is closed under finite convex combinations and
operations in $\cU^\infty_{\rm alg}$. 
\end{Proposition}
\subsection{The energy density}
Let $\om \in \cS_{\sf qH}$ and define $\cF_{\om}$ to consist of
all linear functionals (not, in general, states) 
$\ell$ on $\mathcal{M}_{\om_0} \supset \cA$
given by
\begin{equation} \label{defell}
\ell(B) = \ip{\psi}{\beta_{\om\om_0}(B)\varphi}\,, \qquad B\in
\mathcal{M}_{\om_0} \,,
\end{equation}
for some $\psi,\varphi\in\Had(\om)$. We also 
denote by $\cF$ the set of all linear combinations of finitely
many functionals $\ell_i \in \cF_{\om_i}$, $\om_i \in \cS_{\sf qH}$
and---as in Sec.~\ref{sect:absres}---use $\cV$ to denote the vector space of
functionals on $\cA$ generated (as in Eq.~\eqref{eq:Vdef}) by $\cS$ and $\cU^\infty_{\rm alg}$.
In view of Prop.~\ref{Prop:stability} (see also Thm.\ A.1), 
$\cV$ is necessarily a subset of
$\cF$. Thus,
when investigating properties of the energy density, it is actually enough to
consider elements in $\cF_{\om}$ for arbitrary $\om \in \cS_{\sf qH}$.

Accordingly, let $\om \in \cS_{\sf qH}$ be arbitrarily chosen. Then we
define the one-point function as a linear map from $\cF_{\om}$ to 
$\sD'(M)$ by
\[
\Phi[\ell](f) = \ip{\psi}{\Phi_{\om}(f)\varphi} \,,
\]
for $\ell\in\cF_{\om}$ as in \eqref{defell}, and this is necessarily a
 weak solution to the Klein--Gordon equation.  
Similarly, the two-point function is a weak bisolution defined by
\begin{equation}
\Phi^{\otimes 2}[\ell](f,g) =
\ip{\Phi_{\om}(\overline{f})\psi}{\Phi_{\om}(g)\varphi}  \,,
\label{eq:2pt}
\end{equation}
which satisfies the commutator property
\[
\Phi^{\otimes 2}[\ell](f,g) -\Phi^{\otimes 2}[\ell](g,f)
=iE(f,g)\ell(\II)
\]
as may be seen by a short argument using the Weyl relations and Leibniz'
rule.

The microlocal properties of the one- and two-point functions are easily
determined using the calculus of Prop.~\ref{prop:uloc}. Starting with
the observation that
\[
\ip{\Phi(\overline{f})\psi}{\varphi} = \Phi[\ell](f) =
\ip{\psi}{\Phi(f)\varphi}\,,
\]
for $\psi,\varphi\in\Had(\om)$, the Hadamard condition~(\ref{eq:Hadamard})
and Prop.~\ref{prop:uloc}(ii) imply
\[
(\cN^-)^\dagger\supset\WF(\Phi(\overline{\,\cdot\,})\psi)
\supset \WF(\Phi[\ell]) \subset \WF(\Phi(\,\cdot\,)\varphi)\subset \cN^-
\]
so the one-point function obeys
$\WF(\Phi[\ell])\subset\cN^+\cap\cN^-=\emptyset$ and is therefore
smooth. Turning to the two-point function, Eq.~(\ref{eq:2pt}),
Prop.~\ref{prop:uloc}(iii) and the Hadamard
condition~(\ref{eq:Hadamard}) give
\begin{eqnarray*}
\WF(\Phi^{\otimes 2}[\ell])&\subset& 
\left(\WF(\Phi(\,\cdot\,)\psi)^\dagger\cup\cZ\right)\times
\left(\WF(\Phi(\,\cdot\,)\varphi)\cup\cZ\right)\\
&\subset&(\cN^+\cup\cZ)\times (\cN^-\cup\cZ)\,.
\end{eqnarray*}

In the special case in which $\ell$ is a state, the above inclusion and the
commutator property combine to yield the stronger result that 
$\WF(\Phi^{\otimes 2}[\ell])$ is contained in the right-hand side of 
Eq.~(\ref{eq:RadHad}) and that the two-point function therefore takes the
Hadamard form. By polarisation, it follows that the {\em normal
  ordered two-point function} 
\[
:\Phi^{\otimes 2}:[\ell] = \Phi^{\otimes 2}[\ell]
-\ell(\II)\Phi^{\otimes 2}[\om_0] 
\]
(relative to the reference state $\om_0$ fixed
in Sec.~\ref{sect:dynsys}) can be identified with a smooth function on
$M\times M$ for each $\ell\in\cF$. 
The {\em point-split normal ordered energy density} is defined in terms of
this quantity by 
\begin{equation} \label{defT}
:T:[\ell](x,x') = \frac{1}{2} \left(m^2+\sum_{\mu=0}^s e_\mu^a\nabla_a\otimes
e_\mu^{a'}\nabla_{a'}\right)  :\Phi^{\otimes 2}:[\ell](x,x')
\end{equation}
and is also smooth on $M\times M$; finally,
the {\em normal ordered energy density} itself is given
(cf.~(\ref{eq:cled})) by 
\[
\rrb[\ell](x)= g_{00}(x)^{1/2} :T:[\ell](x,x)\,.
\]
All the quantities defined so far clearly extend to finite linear
combinations of functionals $\ell$ in $\cF_{\om}$ as $\om$ ranges over
$\cS_{\sf qH}$, and hence to $\ell \in \cF$. In particular, $\rrb$ is
defined on $\cS$. As will be proved in Sec.~\ref{sect:dyngen} the
spatial integral of this quantity generates the dynamics.
\begin{Proposition} \label{Prop:dyngen}
For all $t\in\RR$ we have
\begin{equation}
\int_\Sigma d\mu(\ux) \ell([\rrb(t,\ux),A])=
\frac{1}{i}\left.\frac{d}{ds}\ell(\alpha_s A)\right|_{s=0}\,,
\qquad A\in\cU^\infty_{\rm alg},~\ell\in\cV\,,
\label{eq:dyngen}
\end{equation}
where $\cV$ is, as in Sec.~\ref{sect:absres}, the vector space generated
by $\cS$ and $\cU^\infty_{\rm alg}$. 
\end{Proposition}
Eq.~(\ref{eq:dyngen}) clearly
implies that both the assumptions (i) and
(ii) made on the dynamics in Sect.~\ref{sect:absres} are satisfied: to
derive (ii) one simply observes that the right-hand side is
$t$-independent, while (i) follows on replacing $A$ by $\a_t A$. \\

\subsection{The quantum weak energy inequality}

The last step in justifying the structural assumptions of
Sec.~\ref{sect:absres} for our model is the identification of a
state $\om$ obeying a suitable QWEI. For this purpose, $\om$ may be
chosen to be any state in $\cS$ whose 2-point function $\Phi^{\otimes
2}[\om](x,y)$ is invariant under $x\mapsto \tau_t x$, $y\mapsto \tau_t
x$ for any $t\in\RR$.\footnote{Such states certainly exist: for example,
one could use a ground- or KMS-state, but only the invariance and
Hadamard properties are needed below.}

\begin{Proposition} \label{prop:T0}
Let the (unrenormalised) point-split energy density 
$T_0\in\sD'(M\times M)$ be defined by
\[
T_0 = \frac{1}{2} \left(m^2+\sum_{\mu=0}^s e_\mu^a\nabla_a\otimes
e_\mu^{a'}\nabla_{a'}\right)  \Phi^{\otimes 2}[\om]
\]
and define $\Gamma_{\ux}:\RR\to M\times M$ by
$\Gamma_{\ux}(t)=(t,\ux;0,\ux)$. Then:\\ 
i) the pull-back
$\Gamma_{\ux}^*T_0$ exists as an element of  $\sD'(\RR)$ with 
\[
\WF(\Gamma_{\ux}^*T_0)\subset\{(t,\zeta)\mid \zeta>0\}\,;
\]
ii) $\Gamma_{\ux}^*T_0$ is positive-type in the sense that
\begin{equation}
\Gamma_{\ux}^*T_0(\overline{f}\star\widetilde{f})\ge 0 \qquad \hbox{for
all}~f\in\sD(\RR)\,,
\label{eq:pt}
\end{equation}
where $\widetilde{f}(t)=f(-t)$. Furthermore,  $\Gamma_{\ux}^*T_0$ is
a tempered distribution whose Fourier transform is a positive measure
with respect to which $(-\infty,u]$ has finite measure, polynomially bounded
in $u$. 
\end{Proposition}
{\noindent\em Proof:} i) is a direct calculation, using the fact that
$\WF(T_0)$ is contained in $\WF(\Phi^{\otimes 2}[\o])$ which (since all
covectors contained therein are null) has trivial intersection with the
conormal bundle
$$
N_{\Gamma_{\ux}} = \{ ((t,\ux;0,\ux),(0,\uxi;\zeta',\uxi')):
\zeta'\in\RR,~\uxi,\uxi'\in T_{\ux}^*\Sigma\}
$$
of $\Gamma_{\ux}$. Part ii) follows because
$\Gamma_{\ux}^*T_0(f\star\widetilde{g})=\gamma_{\ux}^{(2)*}T_0(f\otimes
g)$ where $\gamma_{\ux}^{(2)}:\RR^2\to M\times M$ is defined by
$\gamma_{\ux}^{(2)}(t,t')=(t,\ux;t',\ux)$. (This map has conormal
bundle $N_{\gamma_{\ux}^{(2)}}=N_{\gamma_{\ux}}\times N_{\gamma_{\ux}}$, where
$$
N_{\gamma_{\ux}} = \{(t,\ux;0,\uxi): t\in\RR,~\uxi\in T_{\ux}^*\Sigma\}
$$
is the conormal bundle for $\gamma_{\ux}:t\mapsto (t,\ux)$. But
$N_{\gamma_{\ux}}$ contains no null covectors, so the pull-back 
$\gamma_{\ux}^{(2)*}T_0$ is well-defined.) Since $T_0$ is positive type
in the sense that $T_0(\overline{F}\otimes F)\ge 0$ for $F\in\sD(M)$, it
follows by Theorem~2.2 in~\cite{AGWQI} that $\gamma_{\ux}^{(2)*}T_0$ is
also positive type in this sense and that~(\ref{eq:pt}) holds. The
remaining statements follow from Theorem~\ref{thm:Boch} in
Sec.~\ref{sect:Boch}, a  
variant of the Bochner-Schwartz theorem. \hfill $\Box$

With the above definitions, the arguments of Sec.~5 of~\cite{AGWQI} may
be adapted straightforwardly\footnote{There are two main differences:
first, a change of parametrisation in the worldline; second,
in~\cite{AGWQI}, the state $\om$ [there denoted $\omega_0$] was additionally
assumed to be a ground state of the time evolution, which has the effect of
limiting the $\zeta$ integration in~(\ref{eq:Qdef}) to $[0,u)$, but is
not otherwise needed 
in the derivation.}
to show that $\om$ obeys a static QWEI in the sense described in
Sec.~2 with respect to the set of states $\cS$, where
$$ 
q(g,\ux) = \int_{\bR} du \, |\widehat{g}(u)|^2 Q(u,\ux) 
$$
and
\begin{equation}
Q(u,\ux) = \frac{1}{2\pi^2}\int_{(-\infty,u)} d\zeta\,\left[
\Gamma_{\ux}^*T_0\right]^\wedge(\zeta)\,.
\label{eq:Qdef}
\end{equation}
In fact, because $\widehat{g}$ is smooth, the static QWEI would be
unchanged if we had instead 
used the integration range $(-\infty,u]$ to define $Q$; however, the
above definition is technically more convenient, as it entails that
$Q(u,\ux)$ is left-continuous in $u$ for each fixed $\ux$. 
We also note that $Q$ is a well-defined nonnegative 
measureable function on $\RR\times\Sigma$ as a consequence of 
Prop.~\ref{prop:T0}(ii); a further consequence of which is that
$Q(u,\ux)$ is polynomially bounded in $u$ for each fixed
$\ux\in\Sigma$. 

The final property required of $Q$ is proved in Sec.~\ref{sect:Q}.
\begin{Proposition} \label{prop:Q} 
${}$ \quad For each $u\in\RR$, $Q(u,\cdot)\in
L^1(\Sigma,d\mu)$; furthermore, 
the function \linebreak
$\fQ(u):=\int_\Sigma d\mu(\ux) Q(u,\ux)$ is
monotonically increasing, left-continuous and polynomially bounded in $u$.
\end{Proposition}
This then implies that $q(g;\,.\,) \in L^1(\Sigma,d\mu)$, thus $\o$
fulfills a static QWEI, but we even have
\begin{Theorem}
The state $\o$ fulfills a limiting static QWEI (with respect to $\cS$), and all
states in $\cS$ fulfill a static QWEI.
\end{Theorem}
{\em Proof:} Note that
\begin{eqnarray}
\frac{1}{\|g_\lambda^2\|_{L^1}}\int_\Sigma d\mu(\ux) q(g_\lambda;\ux)
&=& \frac{1}{\|g^2\|_{L^1}}\int_\Sigma d\mu(\ux) \int_\RR du\,
|\widehat{g}(u)|^2 Q(\lambda u,\ux)\nonumber\\
&=& \frac{1}{\|g^2\|_{L^1}}\int_\RR du\, |\widehat{g}(u)|^2 \fQ(\lambda u)\,,
\label{eq:1LSQ}
\end{eqnarray}
where Fubini's theorem has been used.
Since $\fQ$ is polynomially bounded and $\widehat{g}$ is
of rapid decrease, for any $\epsilon>0$ there exists $U>0$ such that
$$
\int_{|u|>U}  du\, |\widehat{g}(u)|^2\fQ(\lambda u)
<\epsilon \|g^2\|_{L^1}\,, \qquad \lambda\in (0,1)\,.
$$
Thus
\begin{eqnarray*}
\frac{1}{\|g^2\|_{L^1}}\int_\RR du\, |\widehat{g}(u)|^2 \fQ(\lambda u) &\le& 
\frac{1}{\|g^2\|_{L^1}}\int_{|u|<U} du\, |\widehat{g}(u)|^2\fQ(\lambda u)+\epsilon \\
&\le & 2\pi \sup_{|u|< U} \fQ(\lambda u) + \epsilon\\
&= & 2\pi\fQ(\lambda U) + \epsilon
\end{eqnarray*}
for all $\lambda\in(0,1)$,
using monotonicity and left-continuity of $\fQ$. Putting this
together with Eq.~(\ref{eq:1LSQ}) and taking
the limit $\lambda\to 0^+$, we have
$$
\limsup_{\lambda\to 0^+}  \frac{1}{\|g_\lambda^2\|_{L^1}}\int_\Sigma d\mu(\ux)
q(g_\lambda;\ux) \le \limsup_{u\to 0^+} \fQ(u) +
\epsilon = \lim_{u\to 0^+} \fQ(u) + \epsilon\,,
$$
using monotonicity of $\fQ(u)$ again. 
Since $\epsilon$ was
arbitrary, the left-hand side is bounded uniformly in $g$ by
$\fQ(0+):=\lim_{u\to 0^+} \fQ(u)$. Thus $\o$ fulfills a limiting static QWEI
with $0\le \Gamma_\Sigma\le \fQ(0+)$;
the remaining statement follows as indicated in Remark~(i) following
Def.~\ref{D1}. \hfill $\Box$ 

Finally, we strengthen the link between quiescence and ground states as
follows:
\begin{Theorem} \label{Thm:GSQ}
If $\omega\in\cS$ is a non-degenerate ground state with a mass
gap and vanishing one-point function then $\omega$ is quiescent.
\end{Theorem}
{\em Proof:} In the GNS representation
$(\cH_{\om},\pi_{\om},\Omega_{\om})$  induced by
$\omega$,
the dynamics is generated by a self-adjoint operator $H$ with spectrum
$\sigma(H)\subset\{0\}\cup [m_0,\infty)$ for some $m_0>0$ and such that
$0$ is a simple eigenvalue with eigenvector $\Omega$. Thus for any
$F,G\in\sD(M)$, 
\begin{eqnarray*}
\Phi^{\otimes 2}[\omega](F,\tst G) &=&
\ip{\Phi_{\om}(\overline{F})\Omega_{\om}}{e^{iHt}\Phi_{\om}(G)\Omega_{\om}}\\
&=& \int e^{i\zeta t} d\ip{
\Phi_{\om}(\overline{F})\Omega_{\om}}{E_\zeta\Phi_{\om}(G)\Omega_{\om}}\,,
\end{eqnarray*}
where $dE_\zeta$ is the spectral measure for $H$. Due to the spectral
properties of $H$, the nondegeneracy of $\omega$
and the vanishing one-point functions
$\ip{\Omega_{\om}}{\Phi_{\om}(G)\Omega_{\om}}$, 
the (finite) measure
$d\ip{\Phi_{\om}(\overline{F})\Omega_{\om}}{E_\zeta\Phi_{\om}(G)\Omega_{\om}}$ 
is in fact supported in $[m_0,\infty)$.
Similarly, since time translation commutes with $\nabla_a$,
$$
T_0(F,\tst G)  =  \int e^{i\zeta t} d\rho_{F,G}(\zeta)\,,
$$
where $d\rho_{F,G}$ is finite and supported in $[m_0,\infty)$. 

Now take any $f,g\in\sD(\RR)$ and a sequence $\chi_n\to\delta_{\ux}$ in
$\sD_{T^*\Sigma}'(\Sigma)$. Put $F_n=f\otimes \chi_n$, $G_n=g\otimes
\chi_n$. Then for any $h\in\sD(\RR)$, we have
\begin{eqnarray*}
\int \widehat{h}(\zeta)d\rho_{F_n,G_n}(\zeta) &=&
\int dt\, h(t) T_0(F_n,\tst G_n) \\
&=& T_0(F_n,H_n) \\
&\to& \gamma_{\ux}^{(2)*}T_0 (f,h\star g) \\
&=&\Gamma_{\ux}^*T_0(f\star\widetilde{h\star g})\\
&=&(2\pi)^{-1}
\widehat{\Gamma_{\ux}^*T_0}(\widetilde{\widehat{f}}\widehat{g}\widehat{h})\,,
\end{eqnarray*}
where $H_n=[h\star g]\otimes\chi_n$. Furthermore, a similar argument
shows that
$$
\int d\rho_{F_n,G_n}(\zeta) \to (2\pi)^{-1}
\widehat{\Gamma_{\ux}^*T_0}(\widetilde{\widehat{f}}\widehat{g})
$$
which, considering the case $f=\overline{g}$ first and then polarising,
implies
$$
\left| \int \widehat{h}(\zeta)d\rho_{F_n,G_n}(\zeta)\right| \le
C_{f,g}\sup |\widehat{h}|\,.
$$
This provides the necessary uniformity to conclude that
$$
\int \widehat{h}(\zeta)d\rho_{F_n,G_n}(\zeta) \to 
\widehat{\Gamma_{\ux}^*T_0}(\widetilde{\widehat{f}}\widehat{g}\widehat{h})
$$
for all $h\in\Sch(\RR)$. Choosing $h$ so that 
$\supp\widehat{h}\cap[m_0,\infty)=\emptyset$,
and using the fact that linear combinations of functions of the form
$\widetilde{\widehat{f}}\widehat{g}$ are dense in $\Sch(\RR)$, we deduce that
$\widehat{\Gamma_{\ux}^*T_0}$ is supported in $[m_0,\infty)$.

It follows from this that $Q(u,\ux)$ (for each $\ux$) and $\fQ(u)$ are
supported in $[m_0,\infty)$. Since we already know that $\omega$ fulfills a limiting static QWEI with
$0\le \Gamma_\Sigma\le \fQ(0+)$ we may now conclude
that $\Gamma_\Sigma=0$ and that $\omega$ is therefore quiescent.
\hfill$\Box$

\section{Conclusion}
We have now seen that a $C^*$-dynamical system
$(\cA,\{\a_t\}_{t\in\bR})$, together with a suitable class of states
and an energy-density can be constructed for the free scalar field on
a static, spatially compact globally hyperbolic spacetime such that
the assumptions relevant to Sec.~\ref{sect:absres} are
fulfilled. Consequently, we 
obtain from Thm.\ 2.3 that passive states $\omega^p$ exist for the
dynamical system $(\cA,\{\a_t\}_{t\in\bR})$. It may be worth
mentioning again that $\cA$ does not coincide with the Weyl-algebra
$\fA[S,\sigma]$ for the free scalar field on the given static
spacetime, so it is not a priori clear if $\omega^p$ induces a passive 
state (in a $W^*$-sense) on
$\fA[S,\sigma]$. This would follow if $\omega^p$ were found to be
normal to any quasifree Hadamard state, and this is in fact expected
to hold in view of the results of \cite{SV1}.
Furthermore, as shown by Thm.\ 2.6 and Cor.\ 2.7, $\omega^p$ is normal
when energy-compactness holds which is believed to be generically
fulfilled in quantum field theoretical models relevant to particle
physics.

 It should be noted that this
technical complication does not arise in the case of the free Dirac
field, and we should like to emphasize that our methods apply also in
this case, so that also for the case of the Dirac field on a static,
spatially compact and globally hyperbolic spacetime one would be led
to the conclusion that there is a $C^*$-dynamical system together with
a class of states and an energy density fulfilling the assumptions
made in Sec.~\ref{sect:absres}.

In the course of this work, we have also proved various new results
concerning the free scalar field. In particular, we have demonstrated
that the static QWEI bounds obtained in~\cite{AGWQI} exhibit sufficient
spatial regularity to be integrable. We hope also to have demonstrated
the utility of the reformulation (introduced in~\cite{SVW} and developed
further here) of the Hadamard condition as a 
wave-front set condition on Hilbert space-valued distributions.
We expect both this and the microlocal calculus 
of Banach space-valued distributions given in Prop.~\ref{prop:uloc} to have further
applications. 

The results of Sec.~\ref{sect:qfsb} combine with those of
Sec.~\ref{sect:absres} to
give substance to the connection between the
three conditions of dynamical stability mentioned towards the end of
the Introduction. This connection corroborates the point of view
originally advocated by Ford \cite{FordNegEn} that there should be
local constraints on the amount and duration of negative energy
densities in physical quantum states in order that no violations of
the second law of thermodynamics can build up at macroscopic scales. 
However, we re-emphasise that our results indicate a still deeper
equivalence between natural conditions of dynamical stability at 
microscopic, mesoscopic and macroscopic scales.

Of course the question arises how far such an equivalence may be
extended from the situation studied in the present article so as to
apply to interacting quantum fields in generic curved spacetimes 
without time-symmetries, and
hence be regarded as universal. The main difficulty seems to lie in a
missing counterpart of the concept of passivity when the dynamics of a
system is no longer described by a group of automorphisms with
time-independent generators. A further study of possible
generalizations of the passivity concept applicable to quantum field
theory on generic curved spacetimes and their interconnections to
microlocal spectrum condition and QWEIs appears, in the light of the
present results, to offer an interesting line of investigation.

\appendix

\section{Technical appendix}

\subsection{Stability of $\cS$}\label{Sstability}

\begin{Theorem} \label{Had:inv}
Let $\om$ be a quasifree Hadamard state. Then
$\Had(\om)$ is invariant under the action of any element in 
$\beta_{\om\om_0}(\cB)$ where $\cB$ is the $*$-algebra
finitely generated by Weyl operators, elements
of $\cA^\infty$ and elements of $\cU^\infty$. Furthermore, if
$A\in\beta_{\o\om_0}(\cB)$ 
then $[\Phi_{\om}(\cdot),A]\in\sD'(M,\LL(\HH_{\om}))$ with empty
wave-front set.  
In particular, $A\Omega_{\om}\in\Had(\om)$ for any $A\in
\beta_{\om\om_0}(\cB)$.  
\end{Theorem}
{\noindent\em Proof:} 
We will use the following conventions. We will write $W(f)$ for the
Weyl operators $\pi_{\om}({\sf W}(Ef))$ in the GNS-representation of
$\om$, and $\Phi(f)$ for the corresponding generators $\Phi_{\om}(f)$ (${\rm
  e}^{i\Phi(f)} = W(f)$). Moreover, we identify elements $A \in
\mathcal{M}_{\om_0}$ (via $\beta_{\om\om_0}$) with elements in
$\mathcal{M}_{\om}$, i.e., we write $A$ in place of
$\beta_{\om\om_0}(A)$.

Observing these conventions,
it is sufficient to prove that  $\Had(\om)$ is
invariant under Weyl operators, elements of $\cA^\infty$ or operators of
the form $e^{iA}$ for $A=A^*\in\cA^\infty$ and that the required
wave-front set  
condition holds for the commutators of $\Phi$ with such operators. This
will be accomplished in a series of lemmas, 
starting with the case of Weyl operators.

\begin{Lemma} For each $g\in\sD(M)$ and $h\in\sD(M;\RR)$, we have
$W(h) D(\Phi(g))\subset D(\Phi(g))$ and
\begin{equation}
[\Phi(g), W(h)]\psi = -E(g,h)W(h)\psi\,,\qquad\psi\in D(\Phi(g))
\label{eq:pwcom}
\end{equation}
and $[\Phi(\cdot), W(h)]\in\sD'(M,\LL(\HH_{\om}))$ with empty wave-front set.
\end{Lemma}
{\noindent\em Proof:} The domain property and~(\ref{eq:pwcom}) are
standard. Since Eq.~(\ref{eq:pwcom}) implies\\
 $\|[\Phi(g),
W(h)]\|=|E(g,h)|$, the remaining statements follow immediately from
Prop.~\ref{prop:uloc}(i) and the fact that
$\WF(E(\cdot,h))=\emptyset$. \hfill $\Box$

\begin{Lemma} For each $g\in\sD(M)$ and $A\in\cA^\infty$, we have
$A D(\Phi(g))\subset D(\Phi(g))$ and
\begin{equation} \label{B(g)i}
[\Phi(g), A]\psi = B(g)\psi\,,
\end{equation}
where $B(g)\in\cA^\infty$ is given by
\begin{equation} \label{B(g)ii}
B(g) = \int  dt\,f(t) E(\tst h,g)W(\tst h)\,.
\end{equation}
Furthermore, $B(\cdot)\in\sD'(M,\LL(\HH_{\om}))$ and 
$\WF(B(\cdot))=\emptyset$. 
\end{Lemma}
{\noindent\em Proof:} Suppose that $g$ is real-valued and that $A$ is of
the form $\alpha_f W(h)$ [the extension to finite linear combinations
and products of such operators is immediate]. To establish the 
domain property, it suffices to show that
$s\mapsto s^{-1}\|(W(sg)-\II)A\psi\|$ is bounded for $\psi\in D(\Phi(g))$.
Now if $X$ is any bounded operator one may check straightforwardly that 
\begin{equation}
\|XA\psi\| \le 
\int dt\, |f(t)|\, \|X\alpha_t(W(h))\psi\|\,.
\end{equation}
Since $(W(sg)-\II)W(h)=W(h) \left[e^{-isE(g,h)}W(sg)-\II\right]$,
we therefore have
\begin{eqnarray}
s^{-1}\| (W(sg)-\II)A\psi\| &\le& 
s^{-1}\int dt\, |f(t)|\,\| (e^{-isE(g,\tst h)}W(sg)-\II)\psi\| \nonumber\\
&\le & s^{-1}\int dt\, |f(t)|\,\left\{
\| (W(sg)-\II)\psi\| + | e^{-isE(g,\tst h)} -1 |\,\|\psi\| \right\} \nonumber\\
&\le & \|f\|_1 \left\{ \|\Phi(g)\psi\| + 
\sup_{t\in\supp f} | E(g,\tst h) |\,\|\psi\|\right\}\,.
\end{eqnarray}
(Note that the supremum exists as $t\mapsto E(g,\tst h)$ is
continuous.) Thus we have shown that $A\psi\in D(\Phi(g))$ for any
real-valued $g$: since in general $D(\Phi(g)) = D(\Phi(\Re g))\cap
D(\Phi(\Im g))$, the result extends immediately to general $g\in\sD(M)$. 

Eqs.~(\ref{B(g)i}) and~(\ref{B(g)ii}) are now easily checked
using the identity
\begin{equation}
W(f)^* \Phi(g) W(f) = \Phi(g) + E(f,g) \II
\end{equation}
which holds on $D(\Phi(g))$ ($g\in\sD(M)$). 

Next, define $K(g)\in\Coin(\RR)$ by 
\begin{equation}
K(g)(t) = f(t) E(\tst h,g)\,.
\label{eq:Kdef}
\end{equation} 
\begin{Lemma} \label{Lem:WFK}
$K(\cdot)\in\sD'(M,L^1(\RR))$ with $\WF(K)=\emptyset$.
\end{Lemma}
{\noindent\em Proof:} Note first that
\begin{equation}
\|K(g)\|_{L^1}\le \|f\|_{L^1} \sup_{t\in\supp f} \left|E(\tst h,g)\right|
\le C \sup_x |g(x)| 
\end{equation}
so $K(\cdot)\in\sD'(M,L^1(\RR))$. For each $(x,k)\in \dot{T}^*M$, choose
$\cO$ and $\phi$ obeying condition~(A) of Sec.~\ref{sect:uloc} and 
define coordinates on $\cO$ via $y^\mu=\langle
\zeta^\mu,\phi(y)\rangle$ where $\zeta^\mu$ is an arbitrary fixed basis of
$T_x^*M$. Then for each $\chi\in\Coin(\cO)$, $E(\tst h,\chi
e^{i\lambda\langle\eta,\phi\rangle})$ is the usual Fourier
transform\footnote{See footnote~\ref{fn:Fconv}.} at
$\lambda\ell_\mu$ (where $T^*_xM\owns\ell=\ell_\mu\zeta^\mu$) of $F_t=-|\det\gb|^{1/2}
(\chi E\tst h)\circ\kappa^{-1}$ where $\kappa:y\mapsto y^\mu$ is the
coordinate map. Since $F_t\in\Coin(\kappa(\cO))$ with derivatives
varying continuously in $t$, and since $\supp f$ is compact, for any
relatively compact neigbourhood $V$ of $k$ in $T_p^*M$ there exist
constants $C_N$ such that 
\begin{equation}
\sup_{\ell\in V}|E(\tst h,\chi e^{i\lambda\langle\ell,\phi\rangle})|
\le \frac{C_N}{1+\lambda^N}\,,
\qquad \lambda\in\RR^+,~t\in\supp f\,,
\end{equation}
for each $N=1,2,\ldots$.
Accordingly, each $(x,k)\in \dot{T}^*M$ is a regular directed point for $K$, so 
$\WF(K)=\emptyset$.  \hfill $\Box$

It now follows by Prop.~\ref{prop:uloc}(i) and the bound
$\|B(g)\|\le \|K(g)\|_{L^1}$ obtained from Eq.~(\ref{B(g)ii}) that
$B(\cdot)\in\sD'(M,\LL(\HH_{\om}))$ and $\WF(B(\cdot))=\emptyset$.
\hfill $\Box$ 

\begin{Lemma}  For each $g\in\sD(M)$ and $A\in\cA^\infty$, we have
$e^{iA}D(\Phi(g))\subset D(\Phi(g))$ and
\begin{equation}
[\Phi(g), e^{iA}]\psi = C(g)\psi\,,\qquad \psi\in
D(\Phi(g))\,,
\end{equation}
where $C(g)$ is defined by the norm convergent series
\begin{equation}
C(g)=\sum_{k=1}^\infty \frac{i^k}{k!} \sum_{j=0}^{k-1} A^j B(g)A^{k-j-1}\,;
\label{eq:Cform}
\end{equation}
we have $C(\cdot)\in\sD'(M;\LL(\HH_{\om}))$ 
and $\WF(C(\cdot))=\emptyset$. 
\end{Lemma}
{\noindent\em Proof:} By the previous lemma, we may deduce
\begin{equation}
\Phi(g)A^k\psi = A^k\Phi(g)\psi + \sum_{j=0}^{k-1} A^j B(g) A^{k-j-1} 
\end{equation}
for $A\in\cA^\infty$, $k\ge 1$ and $\psi\in D(\Phi(g))$. Norm
convergence in~(\ref{eq:Cform}) follows from the observation
\begin{equation}
\sum_{k=1}^\infty \frac{1}{k!} \sum_{j=0}^{k-1} \|A\|^{k-1}\|B\| =
e^{\|A\|}\|B(g)\|<\infty\,.
\end{equation}
Moreover, the same argument shows that 
$\| C(g)\psi\| \le e^{\|A\|}\|B(g)\| \|\psi\|$. By
Prop.~\ref{prop:uloc}(i), we now conclude that 
$C\in\sD'(M;\LL(\HH_{\om}))$ with $\WF(C(\cdot))=\emptyset$
since $B$ also enjoys these properties. \hfill $\Box$

The proof of Theorem~\ref{Had:inv} is complete. \hfill $\Box$
\\[6pt]
{\noindent\em Proof of Proposition~\ref{Prop:stability}:}
The property $\cS_{\sf qH} \subset \cS$ is obvious since for each $\om
\in \cS_{\sf qH}$ the GNS-vector $\Omega_{\om}$ is contained in
$\Had(\om)$. Also, $\cS$ is closed under finite convex combinations by
its very definition. Now $\cU^\infty_{\rm alg}$ is a sub-$*$-algebra
of $\cB$, and therefore its action on $\Had(\om)$ preserves $\Had(\om)$.
As $\om \in \cS_{\sf qH}$ was arbitrary, this implies immediately that
$\cS$ is preserved under operations in $\cU^\infty_{\rm alg}$.
  \hfill $\Box$

\subsection{The integrated energy density as the dynamical generator}
\label{sect:dyngen}

Let $\gamma_t:\Sigma\to M\times M$ be defined by $\gamma_t(\ux)=
(t,\ux;t,\ux)$. 

\begin{Proposition} \label{prop:dgen}
Let $\om \in \cS_{\sf qH}$ and let $\ell \in \cF_{\om}$. Then
for all $t_0\in\RR$ we have
\begin{equation}
\left(\gamma_{t_0}^*\ell([:T:,A])\right)(g_{00}^{1/2}) =
\frac{1}{i}\left.\frac{d}{ds} 
\ell(\alpha_s A)\right|_{s=0}
\label{eq:dynobj2}
\end{equation}
for any $A\in \beta_{\om\om_0}(\cB)$.
\end{Proposition}
Given this, Prop.~\ref{Prop:dyngen} follows easily.\\[6pt]
{\em Proof of Proposition~\ref{Prop:dyngen}:} Noting that
$\cU^\infty_{\rm alg}\subset \cB$ and that any $\ell\in\cV$ is a finite
convex combination of functionals $\ell_i \in \cF_{\om_i}$, $\om_i \in
\cS_{\sf qH}$,
Eq.~(\ref{eq:dynobj2}) certainly holds for all $A\in\cU^\infty_{\rm
alg}$ and $\ell\in\cV$. To complete the proof, we observe that since
$\ell([:T:,A])$ is smooth we have
\[
\left(\gamma_{t_0}^*\ell([:T:,A])\right)(g_{00}^{1/2}) =\int d\mu(\ux)\,\ell([\rrb(t_0,\ux),A])
\]
and the result follows. \hfill $\Box$
\\[6pt]
{\noindent\em Proof of Prop.~\ref{prop:dgen}:} In the following
subsections, we first establish
Eq.~(\ref{eq:dynobj2}) for Weyl operators and then extend to
$\cA^\infty$ and $\cU^{\infty}_{\rm alg}$. As in A.1, we work in the
GNS representation of $\om$, and write $W(f)$ for $\pi_{\om}({\sf
  W}(Ef))$, $\Phi(f)$ for $\Phi_{\om}(f)$, and we identify $A \in
\mathcal{M}_{\om_0}$ with $\beta_{\om\om_0}(A) \in \mathcal{M}_{\om}$
without writing the $\beta_{\om\om_0}$.

\subsubsection{Weyl operators}

Let us first observe that, for any $A\in\cB$, we have
\[
\ell([:\Phi^{\otimes 2}:,A])(f,g) =
\ip{\Phi(\overline{f})\psi}{[\Phi(g),A]\varphi}
+\ip{[\Phi(\overline{f}),A^*]\psi}{\Phi(g)\varphi}\,,
\]
as may be seen by a straightforward calculation. In the particular case
$A=W(h)$ for $h\in\sD(M;\RR)$, this gives
\begin{equation}
\ell([:\Phi^{\otimes 2}:,W(h)])(x,y) = -
\Phi[\ell_{W(h)}](x) Eh(y) - Eh(x)\Phi[{}_{W(h)}\ell](y)\,,
\label{eq:p2W}
\end{equation}
and, noting that each term in~(\ref{eq:p2W}) is a product of smooth Klein--Gordon
solutions, we use the following lemma.

\begin{Lemma} Suppose $u$ and $v$ are $C^\infty$ solutions to the
Klein--Gordon equation on $(M,\gb)$ and define
\[
\rho(x) = \frac{1}{2}g_{00}(x)^{1/2}\left(\sum_{\mu=0}^s
(e_\mu^a\nabla_a u)|_x (e_\mu^a\nabla_a v)|_x +m^2
u(x)v(x)\right)\,.
\]
Then for any $t_0\in\RR$,
\[
\int_\Sigma d\mu(\ux) \rho(t_0,\ux)=\frac{1}{2}\sigma(\xi^a\nabla_a u,v)\,,
\]
and in the particular case $u=Eh$,
\[
\int_\Sigma d\mu(\ux) \rho(t_0,\ux)=
\frac{1}{2}\int_M d{\rm vol}_\gb(x) v(x) (\xi^a\nabla_a h)(x)\,.
\]
\end{Lemma}
{\em Proof:} We have
\begin{eqnarray*}
\sigma(\xi^a\nabla_a u,v)&=&\int_\Sigma d\mu(\ux) g_{00}^{1/2}\left[
(e_0^a\nabla_a u)( e_0^b\nabla_b v) - v(e_0^a\nabla_a)^2 u\right]
\nonumber\\
&=& \int_\Sigma d\mu(\ux) g_{00}^{1/2}\left[
(e_0^a\nabla_a u)(e_0^b\nabla_b v)- \delta^{ij}v\nabla_a e^a_i
e^b_j\nabla_b u
+m^2uv\right]\nonumber\\
&=& 2\int_\Sigma d\mu(\ux)\rho(t,\ux)\,,
\end{eqnarray*}
where in the second step we have used the Klein--Gordon equation and in
the third step, Gauss' theorem. 
Applying this in the particular case where $u=Eh$ for
$h\in\sD(M)$ and using the fact that $\xi^a\nabla_a$ commutes with $E$, 
we obtain
\[
2\int_\Sigma d\mu(\ux) \rho(t,\ux)=\sigma(\xi^a\nabla_a Eh,v)=
\sigma(E\xi^a\nabla_a h,v)=\int_M d{\rm vol}_\gb(x) v(x) (\xi^a\nabla_a
h)(x)
\]
as required. \hfill $\Box$

Using this result and~(\ref{eq:p2W}) we have
\begin{eqnarray*}
\int_\Sigma d\mu(\ux) \ell([\rrb(t,\ux),W(h)]) &=& 
-\frac{1}{2}\Phi[\ell_{W(h)}+{}_{W(h)}\ell](\xi^a\nabla_a h) \\
&=& -\ip{\psi}{\frac{1}{2}\{\Phi(\xi^a\nabla_a h),W(h)\}\varphi}\\
&=& \frac{1}{i}\left.\frac{d}{ds}\ell(\alpha_s W(h))\right|_{s=0}\,.
\end{eqnarray*}
Here we have used
\begin{Proposition}
\label{prop:c1prop} 
If $\psi,\varphi\in\Had(\om)$ then $s\mapsto \ip{\psi}{\alpha_s
W(f)\varphi}$ is continuously differentiable with derivative
\begin{equation} \label{eq:Wfdiff}
\frac{d}{ds}\ip{\psi}{\alpha_s W(f)\varphi} = -i
\ip{\psi}{\frac{1}{2}\{\Phi(\xi^a\nabla_a \tss f),\alpha_s W(f)\}\varphi}\,.
\end{equation}
\end{Proposition}
{\em Proof:} It is enough to establish differentiability and the form of
the derivative for $s=0$. We begin with two observations. First, if
$\varphi\in\Had(\om)$ and 
$f\to 0$ in $\sD(M;\RR)$ then $W(f)\varphi\to\varphi$ since
\[
\|W(f)\varphi-\varphi\| \le \sup_{t\in\RR\backslash\{0\}} t^{-1}\|
(W(tf)-\II)\varphi\| 
= \|\Phi(f)\varphi\|  \to  0
\]
using the fact that $\Phi(\cdot)\varphi\in\sD'(M,\HH_{\om})$. Second, 
\[
\tss f = f - s\xi^a\nabla_a f +R(s)\,,
\]
where $s^{-1}R(s)\to 0$ in $\sD(M)$ as $s\to 0$. 

Using the Weyl relations, 
\begin{eqnarray*}
s^{-1}\left(W(\tss f)-W(f)\right)\varphi &=& 
s^{-1}\left(e^{iE(\tss f,f)/2}-1\right) W(\tss f-f) W(f)\varphi\nonumber\\
&&+ s^{-1}[W(\tss f-f)-\II]W(f)\varphi 
\end{eqnarray*}
and if $\varphi\in\Had(\om)$ we may use the above observations and the
invariance of 
$\Had(\om)$ under Weyl operators to conclude that the first term on the
right-hand side converges to $-(i/2)E(\xi^a\nabla_a f,f)\varphi$ as $s\to
0$. A further application of the Weyl relations allows us to rewrite the second
term in the form
\begin{eqnarray}
\label{eq:twoterms}
s^{-1}[W(\tss f-f)-\II]W(f)\varphi &=&
s^{-1}e^{-is \nu}W(R(s))
[W(-s\xi^a\nabla_a f)-\II]W(f)\varphi
\nonumber\\
&&+  s^{-1}( W(R(s)e^{-is \nu}-\II)W(f)\varphi\,,
\end{eqnarray}
where $\nu=E(R(s),\xi^a\nabla_a f)/2$.
Estimating the second term of Eq.~(\ref{eq:twoterms}), we see that 
\begin{eqnarray*}
s^{-1}\|( W(R(s))e^{-is \nu}-\II)W(f)\varphi \|
&\le& s^{-1}\|(W(R(s))-\II)W(f)\varphi\| +s^{-1}|e^{-is \nu}-1|\,\|\varphi\|
\nonumber\\
&\le& \| \Phi(s^{-1}R(s))W(f)\varphi\| + \frac{1}{2}|E(R(s),\xi^a\nabla_a
f)|\,\|\varphi\|
\end{eqnarray*}
which tends to zero as $s\to0$ by our two observations. The first term
of Eq.~(\ref{eq:twoterms}) approaches a finite limit as $s\to 0$:
\begin{eqnarray*}
\lefteqn{\ip{\psi}{W(R(s))
s^{-1}[W(-s\xi^a\nabla_a f)-\II]W(f)\varphi}}\nonumber\\
 &\qquad\qquad\qquad\qquad\qquad\qquad=& \ip{W(-R(s))\psi}{
s^{-1}[W(-s\xi^a\nabla_a f)-\II]W(f)\varphi}\nonumber\\
&\qquad\qquad\qquad\qquad\qquad\qquad\to& -i\ip{\psi}{\Phi(\xi^a\nabla_a
f)W(f)\varphi}\,,
\end{eqnarray*}
so we have established that
\[
\left.\frac{d}{ds}\ip{\psi}{\alpha_s W(f)\varphi}\right|_{s=0} = -i
\ip{\psi}{\Phi(\xi^a\nabla_a f)W(f)\varphi}
-\frac{i}{2}E(\xi^a\nabla_a f,f)\ip{\psi}{\varphi}\,,
\]
which can be put into the required form using~(\ref{eq:pwcom}), which
is valid on 
$\Had(\om)$. 

The proof is completed by noting that $s\mapsto\alpha_s W(f)\eta=W(\tss f)\eta$
and $s\mapsto \Phi(\xi^a\nabla_a \tss f)\eta$ are continuous for
$\eta\in\Had(\om)$ (in the first case owing to our first observation and in
the second because $\Phi(\cdot)\eta\in \DD'(M,\HH_{\om})$ and
$s\mapsto \xi^a\nabla_a \tss f$ is continuous from $\RR\to\DD(M)$). Thus
the right-hand
side of~(\ref{eq:Wfdiff}) is continuous in $s$. \hfill $\Box$

\subsubsection{The case $A\in\cA^\infty$}

To extend Eq.~(\ref{eq:dynobj2}) to the case $A\in\cA^\infty$ it
suffices (by Theorem~\ref{Had:inv}) 
to consider $A=\alpha_f W(h)$ for $f\in\Coin(0,\infty)$,
$h\in\sD(M;\RR)$. Now
\begin{eqnarray*}
\frac{1}{i}\left.\frac{d}{ds}
\ell(\alpha_s A)\right|_{s=0}
&=& -i\ell(\delta(\alpha_f W(h))) = -i\ell(\alpha_{-\dot{f}} W(h)) \\
&=& i\int dt\, \dot{f}(t)\ell(\alpha_t W(h))\,.
\end{eqnarray*}
Owing to Prop.~\ref{prop:c1prop}, we may integrate by parts to find
\[
\frac{1}{i}\left.\frac{d}{ds}
\ell(\alpha_s A)\right|_{s=0} = \frac{1}{i}
\int dt\, f(t)\left(\gamma_{t_0}^*\ell([:T:,\alpha_t
W(h)])\right)(g_{00}^{1/2})
\]
using Eq.~(\ref{eq:dynobj2}) for $\alpha_t W(h)$. 

Our goal is now to show that
\begin{equation}
\int dt\, f(t)\left(\gamma_{t_0}^*\ell([:T:,\alpha_t
W(h)])\right)(g_{00}^{1/2}) =
\left(\gamma_{t_0}^*\ell([:T:,\alpha_f W(h)])\right)(g_{00}^{1/2})\,.
\label{eq:pbcont}
\end{equation}
To this end, we observe that
\[
\ell([:T:,\alpha_f W(h)])(F,G) = \int dt\, f(t)
\ell([:T:,\alpha_t W(h)])(F,G)\,, \qquad F,G\in\sD(M)
\]
in which the integrand is compactly supported and---by
Prop.~\ref{prop:c1prop}---continuous.
Accordingly, we may approximate the integral by Riemann sums. Defining
\[
T_N(F,G) = \frac{1}{N}\sum_{n\in\bZ} f(n/N)\ell([:T:,\alpha_{n/N}
W(h)])(F,G)\,,
\]
only finitely many terms contribute for any given $N$, so
$T_N\in\sD'(M\times M)$. Since the Riemann sums approximate the
integral, we have $T_N\to T=\ell([:T:,\alpha_f W(h)])$
weakly, but we require the following stronger convergence property, 
whose proof is deferred to the end of this subsection.
\begin{Lemma} \label{Lemma:TNtoT}
$T_N\to T$ in $\sD_V'(M\times M)$, where $V=(\cN^+\times
\cZ)\cup (\cZ\times\cN^-)$.
\end{Lemma}
Now the conormal bundle of the map $\gamma_{t_0}$ is 
\begin{equation}
N_{\gamma_{t_0}}=\{ (t,\ux,\alpha,\uxi;t,\ux,\beta,-\uxi)\mid 
\alpha,\beta\in\RR,~(\ux,\uxi)\in T^*\Sigma\}
\end{equation}
and has trivial intersection with $V$. It therefore follows that 
$\gamma_{t_0}^* T_N\to\gamma_{t_0}^*
T$ in $\sD'_{\gamma_{t_0}^* V}(\Sigma)$ and in particular that
\[
(\gamma_{t_0}^* T_N)(g_{00}^{1/2})\to (\gamma_{t_0}^*T)(g_{00}^{1/2})=
(\gamma_{t_0}^*\ell([:T:,\alpha_f W(h)]))(g_{00}^{1/2})
\,.
\]
But we also have
\begin{eqnarray*}
(\gamma_{t_0}^* T_N)(g_{00}^{1/2}) &=& \frac{1}{N}\sum_{n\in\bZ} f(n/N)
\left(\gamma_{t_0}^*\ell([:T:,\alpha_{n/N} W(h)]\right)(g_{00}^{1/2})\\
&\longrightarrow & \int dt\, f(t) 
\left(\gamma_{t_0}^*\ell([:T:,\alpha_{t} W(h)]\right)(g_{00}^{1/2})
\end{eqnarray*}
since the final integrand is continuous. Thus~(\ref{eq:pbcont}) is
established and the proof is complete. \hfill$\Box$

{\noindent\em Proof of Lemma~\ref{Lemma:TNtoT}:} It suffices to show
that $U_N\to U$ in $\sD_V'(M\times M)$, where $U_N$ and $U$ are defined
by analogy with $T_N$ and $T$, but with $:\Phi^{\otimes 2}:$ replacing
$:T:$. It is clear that $U_N\to U$ weakly, so we need only show that
$U_N$ converges in $\sD_V'(M\times M)$. Now
\begin{eqnarray*}
\ell([:\Phi^{\otimes 2}:,\alpha_t W(h)])(F,G) &=& 
E(\tst h,F)\ip{\psi}{W(\tst h)\Phi(G)\varphi} +
E(\tst h,G)\ip{\psi}{\Phi(F)W(\tst h)\varphi}
\end{eqnarray*}
and since the two terms each define a smooth distribution on $M\times
M$, their contributions to $U_N$ may be treated separately. Accordingly, we let
\[
U_N^{(1)}(F,G) = \frac{1}{N}\sum_{n\in\bZ} f(n/N) E(\tau_{n/N*}h,F)
\ip{\psi}{W(\tau_{n/N*} h)\Phi(G)\varphi}\,,
\]
and observe that
\[
|U_N^{(1)}(F,G)|\le \frac{1}{N}\sum_{n\in\bZ}
|K(F)(n/N)|\,\|\Phi(G)\varphi\|\,,
\]
where $K(\cdot)\in\DD'(M,L^1(\RR))$ was defined in Eq.~(\ref{eq:Kdef}).
Since $t\mapsto K(F)(t)$ is smooth and compactly supported, we therefore
have
\[
|U_N^{(1)}(F,G)|\le 2 \|K(F)\|_{L^1}\|\Phi(G)\varphi\|
\]
for all sufficiently large $N$. Putting this estimate together 
with the Hadamard condition, Lemma~\ref{Lem:WFK} and
Prop.~\ref{prop:Hpsdom}, we conclude that $U_N^{(1)}$ converges 
in $\sD'_{\cZ\times\cN^-}(M\times M)$. An analogous argument applied to
\[
U_N^{(2)}(F,G) = \frac{1}{N}\sum_{n\in\bZ} f(n/N) E(\tau_{n/N*}h,G)
\ip{\psi}{\Phi(F)W(\tau_{n/N*} h)\varphi}\,,
\]
shows that $U_N^{(2)}$ converges in $\sD'_{\cN^+\times\cZ}(M\times M)$.
Thus $U_N=U_N^{(1)}+U_N^{(2)}$ converges in $\sD_V'(M\times M)$
as required. \hfill $\Box$

\subsubsection{The case $A\in\cU^{\infty}_{\rm alg}$}

The last step in the proof of Prop.~\ref{prop:dgen} is to extend the result
to show that the energy density generates the dynamics for operators in
$\cU^\infty_{\rm alg}$. Here, it is enough to establish
Eq.~(\ref{eq:dynobj2}) with $A$ replaced by an 
operator of the form $e^{iA}$ for $A=A^*\in \cA^\infty$, which may be
arbitrarily well approximated in the graph norm of $\delta$ by the
sequence
\[
S_N = \sum_{n=0}^N \frac{i^n}{n!} A^n
\]
of partial sums. Since $S_N\in \cA^\infty$, and
\[
\left.\frac{1}{i}\frac{d}{ds}\ell(\alpha_s S_N)\right|_{s=0} \longrightarrow 
\left.\frac{1}{i}\frac{d}{ds}\ell(\alpha_s e^{iA})\right|_{s=0}
\]
it remains only to show that
\[
\left(\gamma_{t_0}^*\ell([:T:,S_N])\right)(g_{00}^{1/2})
\longrightarrow
\left(\gamma_{t_0}^*\ell([:T:,e^{iA}])\right)(g_{00}^{1/2})\,.
\]

\begin{Lemma} Suppose $A\in\cA^\infty$. Then
\begin{eqnarray*}
|\ell(A^j[:\Phi^{\otimes 2}:,A]A^k)(f,g)&\le& 
\|A\|^{j+k}\left[\|B(g)\|\,\|\Phi(\overline{f})\psi\| +
\|B(f)\|\,\|\Phi(g)\varphi\|\right]\nonumber\\
&&+(j+k)\|A\|^{j+k-1}\|B(f)\|\,\|B(g)\|
\end{eqnarray*}
for any $j,k\in\NN_0$. 
\end{Lemma}
{\noindent\em Proof:} We have
\[
\ell(A^j[:\Phi^{\otimes 2}:,A]A^k)(f,g) =
\ip{\Phi(\overline{f})A^{*j} \psi}{B(g)A^k\varphi}-
\ip{B(f)^*A^{*j}\psi}{\Phi(g)A^k\varphi}\,.
\]
Now
\[
\Phi(g)A^k\varphi = A^k\Phi(g)\varphi+\sum_{r=0}^{k-1}
A^{r}B(g)A^{k-1-r}\varphi
\]
so
\[
\|\Phi(g)A^k\varphi\| \le
\|A\|^k\|\Phi(g)\varphi\|
+ k\|A\|^{k-1}\|B(g)\|\,.
\]
Putting this 
together with the analogous estimate on $\|\Phi(\overline{f})A^{*j}
\psi\|$, the result is proved by Cauchy-Schwarz.
${}$\hfill $\Box$

In consequence, and using
\[
\ell([:\Phi^{\otimes 2}:,A^n]) = \sum_{j=0}^n 
\ell(A^j[:\Phi^{\otimes 2}:,A]A^{n-1-k})
\]
we obtain
\begin{eqnarray*}
|\ell([:\Phi^{\otimes 2}:,S_N])(f,g)|&\le& 
C\left[\|B(g)\|\,\|\Phi(\overline{f})\psi\|+
\|B(f)\|\,\|\Phi(g)\varphi\|\right]\nonumber\\
&&+C'\|B(f)\|\,\|B(g)\| 
\end{eqnarray*}
for constants $C,C'$ independent of $N$, $f$ and $g$. Applying the
remark following Prop.~\ref{prop:Hpsdom}, the Hadamard condition and the fact that
$B(\cdot)$ has empty wave-front set, we see that
$\ell([:\Phi^{\otimes
2}:,S_N])\to \ell([:\Phi^{\otimes 2}:,e^{iA}])$ in 
$\sD'_V(M\times M)$ as required. 
Proposition~\ref{prop:dgen} is thus proved. \hfill $\Box$

\subsection{A variation on the Bochner-Schwartz Theorem}
\label{sect:Boch}

\begin{Theorem}\label{thm:Boch}
Suppose $S\in\sD'(\RR)$ is of positive type with
\begin{equation}
\WF(S)\subset\{(\tau,\zeta)\mid \zeta>0\}\,.
\label{eq:WFS}
\end{equation}
Then i)~$S\in\Sch'(\RR)$ and $\widehat{S}$ is a polynomially bounded positive
measure; ii) $(-\infty,u]$ has finite (necessarily polynomially bounded)
measure with respect to $\widehat{S}$; iii) if $f_n$ is any sequence of
Schwartz test functions with $f_n(\zeta)$ monotonically increasing to
$\chi_{(-\infty,u)}(\zeta)$ for each $\zeta\in\RR$ then $\widehat{S}(f_n)$
is monotonically increasing and
\begin{equation}
\lim_{n\to\infty} \widehat{S}(f_n) = \int_{(-\infty,u)}
d\zeta\,\widehat{S}(\zeta)\,. 
\end{equation}
\end{Theorem}
{\noindent\em Proof:} Part (i) is the usual Bochner-Schwartz theorem
(Theorem~IX.10 in~\cite{RS2}), 
while (iii) follows by the monotone convergence theorem if (ii) holds. 
It is enough to prove (ii) for $u=0$. To this end, let $f\in C^\infty(\RR)$ be
nonnegative with $\supp f\subset(-\infty,1)$ and $f=1$ on $\RR^-$. 
Decomposing $f$ as 
$f(\zeta)=\sum_{n=0}^\infty g(\zeta+n)$ where
$g\in\Coin(-1,1)$ is nonnegative, we claim that 
\begin{equation}
\int d\zeta\,\widehat{S}(\zeta) g(\zeta-\eta) \to 0 
\label{eq:decclm}
\end{equation}
rapidly as $\eta\to -\infty$. Thus
\begin{equation}
\sum_{n=0}^\infty \int d\zeta\,\widehat{S}(\zeta) g(\zeta+n) <\infty
\end{equation}
and since each term in this series is positive, the monotone convergence
theorem entails that $\int
d\zeta\,\widehat{S}(\zeta)f(\zeta)<\infty$. Accordingly, 
$\int_{\RR^-} d\zeta\,\widehat{S}(\zeta)<\infty$ and the result is proved.

It remains to prove our claim~(\ref{eq:decclm}). Let
$G(\zeta)=(\widehat{S}\star\widetilde{g})(\zeta)=\widehat{S}(g(\cdot-\zeta))$, 
where we have written $\widetilde{g}(\zeta)=g(-\zeta)$. Then $G$ is smooth and
polynomially bounded, with polynomially bounded derivatives. Moreover,
by (i), since $g$ is nonnegative, $G$ is also nonnegative and we may
write
\begin{equation}
G(\zeta) = \int d\eta\,\widehat{S}(\eta) g(\eta-\zeta)\,.
\end{equation}
Using the convolution theorem,
\begin{equation}
S(\chi \widetilde{g}^\vee e_\eta) = \int
d\zeta\,G(\zeta)\widehat{\chi}(\eta-\zeta) 
\end{equation}
for any $\chi\in\sD$. Choose $\chi$ so that $\widehat{\chi}$ is
nonnegative and $\widehat{\chi}(\zeta)>1$ for $|\zeta|<1$. Then for any
$\epsilon\in(0,1)$ we have
\begin{equation}
0\le \inf_{\zeta\in(\eta-\epsilon,\eta+\epsilon)} G(\zeta)\le (2\epsilon)^{-1}
S(\chi \widetilde{g}^\vee e_\eta)\,,\qquad\,\eta\in\RR
\end{equation}
and since $G$ has polynomially bounded first derivative, there exists
$C>0$ and $r>0$ such that
\begin{equation}
0 \le G(\eta)\le 
(2 \epsilon)^{-1}
S(\chi {\widetilde{g}}^\vee e_\eta) +\epsilon C(1+|\eta|)^r\,,\qquad
\eta\in\RR,~0<\epsilon<1 \,.
\end{equation}
In particular, taking $\epsilon = (1+|\eta|)^{-(N+r+1)}$ and using
the hypothesis~(\ref{eq:WFS}) on $\WF(S)$, it follows that
$(1+|\eta|)^NG(\eta)\to 0$ as $\eta\to-\infty$ for each $N\ge 0$,
thereby establishing our claim.  \hfill $\Box$

\subsection{Integrability of $Q(u,\cdot)$} \label{sect:Q}

{\noindent\em Proof of Proposition~\ref{prop:Q}:} 
Let $f_n$ be a sequence of Schwartz test functions such that
$\widehat{f_n}$ is a sequence of nonnegative (Schwartz) functions
monotonically increasing to $\chi_{(-\infty,u)}$. Then by
Theorem~\ref{thm:Boch}, 
\begin{equation}
\Gamma_{\ux}^*T_0(\overline{f_n}\star\widetilde{f_n}) = (2\pi)^{-1}
\widehat{\Gamma_{\ux}^*T_0}(|\widehat{f_n}|^2)\to \pi Q(u,\ux)
\end{equation}
for each $\ux\in\Sigma$. Choosing $g_n\in\sD(\RR)$ such that $g_n-f_n\to
0$ in $\Sch(\RR)$ and using the positive type property of
$\Gamma_{\ux}^*T_0$, we have 
\begin{equation}
0\le \Gamma_{\ux}^*T_0(\overline{g_n}\star\widetilde{g_n}) \to \pi
Q(u,\ux)\,,\qquad \ux\in\Sigma\,.
\label{eq:gnconv}
\end{equation}
\begin{Lemma} \label{lem:equiv}
For each $h\in\sD(\RR)$, $\Gamma_{\ux}^*T_0(h)\in
L^1(\Sigma,d\mu(\ux))$ and
\begin{equation}
\int d\mu(\ux) \Gamma_{\ux}^*T_0(h)=\fT(h)\,,
\label{eq:equiv}
\end{equation}
where
\begin{equation}
\fT(h)=\Gamma^*T_0(h\otimes g_{00}^{-1/2})
\end{equation}
and $\Gamma:M\to M\times M$ is given by $\Gamma(t,\ux)=(t,\ux;0,\ux)$. 
The distribution $\fT$ is of positive type, with wave-front set contained in
the right-hand side of~(\ref{eq:WFS}). 
\end{Lemma} 
Applying the Lemma to $h=\overline{g_n}\star\widetilde{g_n}$, each
$\Gamma_{\ux}^*T_0(\overline{g_n}\star\widetilde{g_n})$ is an
$L^1(\Sigma,d\mu(\ux))$ function with norm ($=$ integral) converging to 
\begin{equation}
\lim_{n\to\infty} \int d\mu(\ux)\,\Gamma_{\ux}^*T_0(\overline{g_n}\star\widetilde{g_n})
= \lim_{n\to\infty} \fT(\overline{g_n}\star\widetilde{g_n}) =
\lim_{n\to\infty} \fT(\overline{f_n}\star\widetilde{f_n})
=\int_{(-\infty,u)} d\zeta\,\widehat{\fT}(\zeta)\,,
\end{equation}
where we have again used~Theorem~\ref{thm:Boch}, now applied to $\fT$.
Putting this together with~(\ref{eq:gnconv}) and applying Fatou's
lemma~\cite{RS1}, we conclude that $Q(u,\cdot)\in L^1(\Sigma,d\mu)$ with
\begin{equation}
0\le \int_\Sigma d\mu(\ux) Q(u,\ux) \le 
\frac{1}{\pi}\int_{(-\infty,u)} d\zeta\,\widehat{\fT}(\zeta)\,,
\end{equation}
the right-hand side of which is polynomially bounded by
Theorem~\ref{thm:Boch}. Thus $\fQ(u)=\int_\Sigma d\mu(\ux) Q(u,\ux)$ is
polynomially bounded; monotonicity and left-continuity follow from
the same properties of $Q(\cdot,\ux)$ and the monotone convergence
theorem. \hfill $\Box$

{\noindent\em Proof of Lemma~\ref{lem:equiv}:} 
Note first that $\Gamma_{\ux}^*T_0=\g_{\ux}^*\Gamma^*T_0$,
where $\g_{\ux}:\RR\to M$ is given by $\g_{\ux}(t)=(t,\ux)$.
Now, considering the definition of the distributional pull-back
(cf.\ \cite{Hormander1}, Thms.\ 8.2.10, 8.2.12) we have
\begin{equation}
\Gamma_{\ux}^*T_0(h)=\g_{\ux}^*\Gamma^*T_0(h) =
\left[(h\otimes\delta_{\ux})\Gamma^*T_0\right](1_M)\,.
\label{eq:equiva}
\end{equation}
On the other hand, define $(\Gamma^*T_0)_h\in\sD'(\Sigma)$ by
$(\Gamma^*T_0)_h(H)=\Gamma^*T_0(h\otimes H)$. Then
$\WF((\Gamma^*T_0)_h)=\emptyset$ and so
\begin{equation}
\Gamma^*T_0(h\otimes g_{00}^{-1/2})=\int_\Sigma d\mu(\ux) 
\left((\Gamma^*T_0)_h\delta_{\ux}\right)(1_\Sigma)
\label{eq:equivb}
\end{equation}
[the $g_{00}^{-1/2}$ arises because the preferred
densities on $\RR$, $\Sigma$ and $M=\RR\times\Sigma$ which identify
distributions and distributional densities are related by
$\rho_M(t,\ux)=g_{00}(\ux)^{1/2}\rho_\RR(t)\rho_\Sigma(\ux)$].
Finally, if $\delta_n$ is a sequence in $\sD(\Sigma)$, converging to $\delta_{\ux}$ in
the H\"ormander pseudo-topology on $\sD_{\WF(\delta_{\ux})}'(\Sigma)$ we
may calculate
\begin{equation}
\left((\Gamma^*T_0)_h\delta_{\ux}\right)(1_\Sigma) = \lim_{n\to\infty}
(\Gamma^*T_0)_h(\delta_n) =\lim_{n\to\infty} 
\left((h\otimes f_n) \Gamma^*T_0\right)(1_M)
=\left[(h\otimes\delta_{\ux})\Gamma^*T_0\right](1_M)\,.
\end{equation}
Putting this together with~(\ref{eq:equiva}) and~(\ref{eq:equivb}) we
obtain~(\ref{eq:equiv}).  

It is clear from~(\ref{eq:equiv}) that $\fT$ is of positive type, and a
direct calculation of its wave-front set shows that $\WF(\fT)$ is contained in
the right-hand side of~(\ref{eq:WFS}). \hfill $\Box$

\subsection{Correlation functions for states on $\cA$}
\label{sect:corell}
We begin our discussion of $n$-point correlation function for states
on $\cA$ with the following definitions:

Let $n \in \bN$ and $F \in \mathscr{S}(\bR^n)$, and define
$$ {\bf W}_F^{(\a)}(f_1,\ldots,f_n) = \int dt_1 \cdots dt_n\, F(t_1,\ldots,t_n)
\a_{t_1}(W(f_1)) \cdots \a_{t_n}(W(f_n)) $$
for $f_j \in C_0^\infty(M,\bR)$, where we work in the defining
representation $\pi_{\o_0}$ of $\cA$ with $W(f_j) = \pi_{\o_0}({\sf
  W}(Ef_j))$; hence the expression on the right hand side exists as a
weak integral in $\cB(\cH_{\o_0})$. Next we claim that the just
defined objects are contained in $\cA$. This can be seen by noting
that there are sequences $h^{(k)}_j \in \mathscr{S}(\bR)$, $k \in
\bN$, so that 
$$ \sum_{k = 1}^N h_1^{(k)} \otimes \cdots \otimes h^{(k)}_n
\underset{N \to \infty}{\longrightarrow} F \ \ {\rm in}\ \
\mathscr{S}(\bR^n) \ \ {\rm and} \ \ \sum_{k = 1}^{\infty}
 ||h_1^{(k)} \otimes \cdots \otimes h^{(k)}_n||_{\infty} < \infty\,.$$
This implies that $\a_{h^{(k)}_1}W(f_1)\cdots \a_{h^{(k)}_n}W(f_n)$
converges in norm to ${\bf W}^{(\a)}_F(f_1,\ldots,f_n)$ as $k \to
\infty$, and since clearly each $\a_{h^{(k)}_j}W(f_j)$ is in $\cA$,
${\bf W}^{(\a)}_F(f_1,\ldots,f_n)$ is also contained in $\cA$.

Now recall the definition of the {\it $n$-point correlation functions}
of a state $\o$ on the Weyl-algebra $\fA[S,\sigma]$\footnote{More
  precisely, $\o$ is to be viewed here as a state on the represented
  Weyl-algebra $\pi_{\o_0}(\fA[S,\sigma])$}: One says that $\o$ is
{\it $C^\infty$-regular} if, for each $n \in \bN$, the map
$$ \bR^n \times C_0^\infty(M,\bR)^n \owns
(t_1,\ldots,t_n;f_1,\ldots,f_n) \mapsto \omega(W(t_1f_1)\cdots
W(t_nf_n)) $$
is $C^\infty$ with respect to the $t_j$ and if the
derivatives
$$ w^{(\o)}_n(f_1,\ldots,f_n) = (-i)^n \left. \frac{\partial}{\partial
    t_1} \cdots \frac{\partial}{\partial t_n}\right|_{t_j = 0}
\o(W(t_1f_1) \cdots W(t_nf_n)) $$
induce, by requiring complex-linearity,
distributions $w^{(\o)}_n \in \mathscr{D}'(M^n)$. These distributions
are then called the $n$-point correlation functions of $\o$. As is
well-known
 (essentially
by Wightman's reconstruction theorem, cf.\ \cite{StrWi}), a
$C^\infty$-regular state $\o$ induces a state $\overline{\o}$ on the
algebra $\mathscr{F}$ of abstract Klein-Gordon field operators. This
algebra $\mathscr{F}$ is 
a $*$-algebra
generated by a unit element ${\bf 1}$ and a
family of elements $\boldsymbol{\phi}(f)$, $f \in C_0^\infty(M)$,
subject to the following relations:
\begin{itemize}
\item[(1)] $f \mapsto \boldsymbol{\phi}(f)$ is $\bC$-linear,
\item[(2)] $\boldsymbol{\phi}(f)^* = \boldsymbol{\phi}(\overline{f})$,
\item[(3)] $\boldsymbol{\phi}((g^{ab}\nabla_a\nabla_b + m^2)f) = 0$,
\item[(4)] $[\boldsymbol{\phi}(f_1),\boldsymbol{\phi}(f_2)] = i
  \sigma(Ef_1,Ef_2){\bf 1}$\,. 
\end{itemize}
The state $\overline{\o}$ is then defined on $\mathscr{F}$ by setting
$\overline{\o}(a{\bf 1}) = a$ ($a \in \bC$), 
$$ \overline{\o}(\boldsymbol{\phi}(f_1) \cdots \boldsymbol{\phi}(f_n))
= w^{(\o)}_n(f_1,\ldots,f_n)\,, $$
and by requiring complex linearity. By the properties of the
GNS-representation, the algebra $\mathscr{F}$ is $*$-isomorphic to the
algebra of field operators $\Phi_{\omega}(f) = -i
\left. \frac{d}{dt}\right|_{t=0} \pi_{\o}({\sf W}(tEf))$ by
identifying $\boldsymbol{\phi}(f)$ and $\Phi_{\o}(f)$ (and by
identifying unit operators).

Proceding along these lines, one can define an analogue of $n$-point
correlation functions also for states $\o$ on $\cA$. We will say that 
a state $\o$ on $\cA$ is {\it $C^\infty$-regular} if for each
$n \in \bN$ and all $F \in
\mathscr{S}(\bR^n)$ 
  the map
$$ \bR^n \times C_0^\infty(M,\bR)^n \owns
(t_1,\ldots,t_n;f_1,\ldots,f_n) \mapsto \omega({\bf W}^{(\a)}_F(t_1f_1,\ldots,t_nf_n)) $$
is $C^\infty$ with respect to the $t_j$  and if the
derivatives 
$$ \widetilde{w}^{(\o)}_n(F;f_1,\ldots,f_n) = (-i)^n
\left. \frac{\partial}{\partial
    t_1} \cdots \frac{\partial}{\partial t_n}\right|_{t_j = 0}
\o({\bf W}^{(\a)}_F(t_1f_1,\ldots,t_nf_n)) $$ 
induce distributions $\widetilde{w}^{(\o)}_n \in
(\mathscr{S}(\bR^n) \otimes \mathscr{D}(M^n))'$. 

Notice that there are many vector states with respect to the defining
representation of $\cA$ which are $C^\infty$-regular states on $\cA$:
For example, all states on $\cA$ induced by vectors $W(f)\O_{\o_0}$, $f \in
C_0^\infty(M,\bR)$, have this property.

We now proceed to establish the following result.
\begin{Theorem}
Let $\o$ be a state on $\cA$ which is $C^\infty$-regular. Let $F \in
C_0^{\infty}(\bR^n)$ with $\int dt_1\cdots dt_n\,F(t_1,\ldots,t_n) =
1$ and define $F^{(\l)}(t_1,\ldots,t_n) =
\l^{-n}F(t_1/\l,\ldots,t_n/\l)$ $(\l > 0)$ so that $F^{(\l)}$ approximates the
$n$-dimensional Dirac-distribution as $\l \to 0$.

Then for all $n \in \bN$ the limits
$$ \overline{w}^{(\o)}_n(f_1,\ldots,f_n) = \lim_{\l \to 0}
\widetilde{w}^{(\o)}_n(F^{(\l)};f_1,\ldots,f_n)\,, \quad f_j \in
\mathscr{D}(M)\,,$$
exist, and induce distributions $\overline{w}^{(\o)}_n \in
\mathscr{D}'(M^n)$ which will be called $n$-point correlation
functions of $\o$.

Moreover, the $\overline{w}^{(\o)}_n$, $n \in \bN$, induce a state
$\overline{\o}$ on the algebra $\mathscr{F}$ of field operators upon
setting $\overline{\o}(a{\bf 1}) = a$ ($ a \in \bC$) and 
$$ \overline{\o}(\boldsymbol{\phi}(f_1) \cdots \boldsymbol{\phi}(f_n))
= \overline{w}^{(\o)}_n(f_1,\ldots,f_n)\,, \quad f_j \in
\mathscr{D}(M)\,,$$
and by requiring linearity.
\end{Theorem}
{\it Proof. } Let ${\bf s} = (s_1,\ldots,s_n) \in \bR^n $ and define,
for $F \in \mathscr{S}(\bR^n)$,
$$ F_{\bf s}(t_1,\ldots,t_n) = F(t_1 - s_1,\ldots,t_n - s_n)\,.$$
Owing to the definition of ${\bf W}^{(\a)}_F(f_1,\ldots,f_n)$, it
holds that
$$ {\bf W}^{(\a)}_{F_{\bf s}}(f_1,\ldots,f_n) = {\bf
  W}^{(\a)}_F(\t_{s_1*}f_1,\ldots,\t_{s_n*}f_n)\,,$$
and this implies
$$ \widetilde{w}^{(\o)}_n(F_{\bf s};f_1,\ldots,f_n) =
\widetilde{w}^{(\o)}_n(F;\t_{s_1*}f_1,\ldots,\t_{s_n*}f_n)\,.$$
Let us now denote, for simplicity of notation, the distribution in
$(\mathscr{S}(\bR^n)\otimes \mathscr{D}(M^n))'$ induced by
$\widetilde{w}^{(\o)}_n$ simply by $w$. Then the last equation implies
for this distribution the relation
\begin{equation} \label{equivar}
 w((F * G) \otimes \varphi) = w(F \otimes (G \star \varphi))\,,  \quad
 F \in \mathscr{S}(\bR^n)\,, \ \ G \in \mathscr{D}(\bR^n)\,, \ \
 \varphi \in \mathscr{D}(M^n)\,,
\end{equation}
where $F*G$ is the usual convolution of functions on $\bR^n$ and
$$ (G \star \varphi)(t_1,\ux_1,\ldots,t_n,\ux_n) = \int ds_1\cdots
ds_n \, G(s_1,\ldots,s_n)\varphi(t_1 - s_1,\ux_1,\ldots,t_n -
s_n,\ux_n)\,.$$
Now we define
$$ (\D_{(1)}\varphi)(t_1,\ux_1,\ldots,t_n,\ux_n) =
 -(\partial_{t_1}^2 + \cdots +
 \partial_{t_n}^2)\varphi(t_1,\ux_1,\ldots,t_n,\ux_n) $$
and
$$ (\D G)(t_1,\ldots,t_n) = -(\partial_{t_1}^2 + \cdots +
\partial_{t_n}^2)G(t_1,\ldots,t_n)\,.$$
Then relation \eqref{equivar} implies for all $F \in
\mathscr{S}(\bR^n)$, $G \in \mathscr{D}(\bR^n)$, $\varphi \in
\mathscr{D}(M^n)$ the following chain of equations:
\begin{eqnarray*}
w((F * G) \otimes \varphi) & = & w((1- \D)^{-m}F * (1 - \D)^mG \otimes
\varphi)\\
 & = & w((1- \D)^{-m}F \otimes (1- \D)^mG \star \varphi) \\
 & = & w((1 - \D)^{-m}F \otimes G \star (1- \D_{(1)})^m\varphi) \\
 & = & w((1- \D)^{-m}F * G \otimes (1- \D_{(1)})^m\varphi)
\end{eqnarray*}
which is valid for all $m \in \bN$. Letting $G$ tend to the $n$-dimensional
Dirac-distribution, we find
$$ 
 w(F \otimes \varphi) = w((1- \D)^{-m}F \otimes (1-
 \D_{(1)})^m\varphi)
$$
for all $F \in \mathscr{S}(\bR^n)$, $\varphi \in \mathscr{D}(M^n)$ and
$m \in \bN$. Hence, exploiting the regularizing property of $(1-
\D)^{-m}$, one may choose $m$ so large that 
$$ \lim_{\l \to 0}\, w(F^{(\l)} \otimes \varphi) = \lim_{\l \to 0}\,
w((1 -\D)^{-m}F^{(\l)} \otimes (1- \D_{(1)})^m\varphi)$$
exists (uniformly in $\varphi$) owing to the continuity of the
functional $w$; this then implies that the resulting limit is a
distribution in $\mathscr{D}'(M^n)$ with respect to $\varphi$.

In order to show that the above definition of $\overline{\o}$ really
defines a state on $\mathscr{F}$, one needs to check first that
$\overline{\o}$ induces a linear functional on $\mathscr{F}$, i.e.\
that it respects the relations expressed in $(1)$-(4) above. Relation
(1) is fulfilled since the $\overline{w}^{(\o)}_n$ induce
distributions (and are thus multilinear). The next relation (2) is
essentially a consequence of $[(it)^{-1}(W(tf) - {\bf 1})]^* = (i/t)(W(-tf)
-{\bf 1})$ and so its proof is completely analogous to showing that the
$n$-point correlation functions of a $C^\infty$-regular state on the
Weyl-algebra induce a state on $\mathscr{F}$. Moreover,
$\overline{w}^{(\o)}_n(f_1,\ldots,f_n) = 0$ if any of the $f_j$ is in
the range of the Klein-Gordon operator since, in this case,
$W(t_jf_j) = {\bf 1}$, so that $\overline{\o}$ respects relation
$(3)$. Finally, to check the CCR, note that
$$ {\bf W}^{(\a)}_F(f_1,\ldots,f_j,\ldots,f_k,\ldots,f_n) = {\bf
  W}^{(\a)}_{F \cdot s_{jk}}(f_1,\ldots,f_k,\ldots,f_j,\ldots,f_n) $$ 
with $s_{jk}(t_1,\ldots,t_n) =
{\rm e}^{i\sigma(E\t_{t_j*}f_j,E\t_{t_k*}f_k)}$. Inserting this into the
definition of the $\widetilde{w}^{(\o)}_n$ yields
\begin{eqnarray*}
& & {} \hspace*{-3cm}
\widetilde{w}^{(\o)}_n(F;f_1,\ldots,f_j,\ldots,f_k,\ldots,f_n) -
\widetilde{w}^{(\o)}_n(F;f_1,\ldots,f_k,\ldots,f_j,\ldots,f_n)
 \\
& = &
\widetilde{w}^{(\o)}_{n-2}(G;f_1,\ldots,\overset{{\sf x}}{f}_j,
\ldots,\overset{{\sf x}}{f}_k,\ldots,f_n) 
\end{eqnarray*} where an {\sf x} over a symbol means that the corresponding
entry doesn't appear, and the function $G$ is given by
$$ G(t_1,\ldots,\overset{{\sf x}}{t}_j,\ldots,\overset{{\sf
    x}}{t}_k,\ldots,t_n) = 
 \int
 dt_jdt_k\,F(t_1,\ldots,t_n)i\sigma(E\t_{t_j*}f_j,E\t_{t_k*}f_k)\,.$$
These equations establish that $\overline{\omega}$ respects the CCR.

What remains to be checked is the positivity of $\o$. To this end, let
$$ P = \left(a_0{\bf 1} + \sum_{k = 1}^N a_k\boldsymbol{\phi}(f_1^{(k)})
\cdots \boldsymbol{\phi}(f^{(k)}_{m_k})\right)^*\left(a_0{\bf 1} + \sum_{k = 1}^N a_k\boldsymbol{\phi}(f_1^{(k)})
\cdots \boldsymbol{\phi}(f^{(k)}_{m_k})\right)$$
with $a_0,\ldots,a_N \in \bC$ and $f^{(k)}_j \in C_0^\infty(M,\bR)$ be
a generic positive element of $\mathscr{F}$. Then it holds that
\begin{eqnarray*}
\overline{\o}(P) & = & |a_0|^2 + 2{\rm Re}\, a_0 \sum_{k =
  1}^N\overline{w}^{(\o)}_{m_k}(f^{(k)}_1,\ldots,f^{(k)}_{m_k}) \\
 & & + \ \sum_{k,\ell = 1}^N \overline{a}_ka_{\ell}
 \overline{w}^{(\o)}_{m_k +
   m_\ell}(f_{m_k}^{(k)},\ldots,f_1^{(k)},f_1^{(\ell)}.\ldots,f^{(\ell)}_{m_{\ell}}) \\
& = & \lim_{h \to \delta}\,\lim_{t\to 0}\,\o(Q(h,t)^*Q(h,t))
\end{eqnarray*}
where we define for $h \in C_0^{\infty}(\bR)$ with $\int dt\,h(t) = 1$,
$$ Q(h,t) = a_0{\bf 1} + \sum_{k = 1}^N a_k\left(\frac{\a_hW(tf^{(k)}_1) -
  {\bf 1}}{it}\right) \cdots \left(\frac{\a_hW(tf^{(k)}_{m_k}) -
  {\bf 1}}{it}\right) \,.$$
Hence the expressions over which the limits are taken are
non-negative, and thus $\overline{\o}(P) \ge 0$.${}$\quad \ ${}$ \hfill $\Box$
\\[24pt]
{\bf Acknowledgments }
We would like to thank Bernd Kuckert for discussions related to
passivity, and for comments on the manuscript. We also thank him 
for having made his recent results available to us prior to publication.
We also thank Muharrem K\"usk\"u for drawing our attention to some
typographical errors in the preprint of this paper. 
The work of CJF was assisted by grant NUF-NAL/00075/G from the Nuffield
Foundation and, in its later stages, by EPSRC grant GR/R25019/01.

 
%
\end{document}